\documentclass[final,5p,times, twocolumn]{elsarticle}

\usepackage{amsmath,amssymb,amsfonts}
\usepackage{algorithm}
\usepackage [noend]{algpseudocode}
\usepackage{algpseudocode}
\usepackage{subfigure}
\usepackage{booktabs}
\usepackage{graphicx,color}
\usepackage{lscape}
\usepackage{textcomp}
\usepackage{colortbl}
\usepackage{tcolorbox}
\usepackage{tabularx}
\usepackage{url}
\usepackage{hyperref}
\usepackage{placeins}
\usepackage{array}
\usepackage{float}
\usepackage{multirow}
\usepackage{titlesec}
\usepackage{xcolor}
\usepackage{lineno}
\usepackage{threeparttable}

\titleformat{\paragraph}
{\normalfont\normalsize\itshape}{\theparagraph}{1em}{}
\titlespacing*{\paragraph}
{0pt}{3.25ex plus 1ex minus .2ex}{1.5ex plus .2ex}

\begin{document}

\begin{frontmatter}
	
	
	
	\title{Social Popularity of GitHub Projects: A Lifeline or a Liability?}
	
	
	\author{Mohit Kaushik, Kuljit Kaur Chahal} 
	
	\affiliation{organization={Department of Computer Science, Guru Nanak Dev University},
		addressline={}, 
		city={Amritsar},
		postcode={143005}, 
		state={Punjab},
		country={India}}
	
	\begin{abstract}
		
	Social coding platforms such as GitHub host millions of repositories, yet many suffer from high mortality rates. Despite this, several survival factors remain poorly understood. Human capital is widely recognized as essential. Social attention while often assumed to be a lifeline, can become a liability. Structural features that improve onboarding, such as code readability and documentation, may also accelerate the cessation of active development when combined with massive visibility.
	To examine these dynamics, we analyzed more than 73,000 GitHub repositories using an Accelerated Failure Time (AFT) survival framework, which accounts for the time‑varying nature of predictors. Our study identifies human capital as the most critical determinant of project survival. In contrast, excessive social attention emerges as a liability, and when coupled with accessibility features, it amplifies the risk of project inactivity. Importantly, when the number of contributors interacts with social popularity, the protective effect of labor becomes visible, highlighting the need for governance strategies that balance visibility with labor capacity to ensure the long‑term resilience of open‑source projects.
	\end{abstract}
	
	
\end{frontmatter}

	

\section{Introduction}\label{sec:intro}

In the modern era, open-source software (OSS) has become the backbone of digital infrastructure, as the majority of organizations utilize it to some extent~\cite{synopsys2024ossra}. Major social coding platforms like GitHub host millions of repositories; although the majority are experimental or trivial~\cite{kalliamvakou2014promises}, a significant percentage still suffers from severe abandonment and a short lifespan~\cite{Ait2022}. To understand this high mortality rate, studies continually highlight human capital as the primary lifeline for a project~\cite{Samoladas2010,Robinson2022,park2025analyzing}. Historically, project survival has relied almost entirely on the sustained effort of volunteer contributions, and research has demonstrated that an active core team protects a project from stagnation~\cite{mockus2002two}. With the advent of social coding platforms, community attention also emerges as a crucial survival factor~\cite{avelino2019abandonment,park2025analyzing}.

Naturally, a baseline level of community attention is required for a project to survive by attracting new contributors~\cite{park2024assessing,borges2016understanding}. However, the architecture of modern social coding has introduced a new and challenging volunteer model: hyper-visibility~\cite{dabbish2012social}. While ``mega-projects" such as Linux or Mozilla Firefox survive at a massive scale by adopting commercial-style governance and employing corporate-backed, full-time developers~\cite{mockus2002two,fitzgerald2006transformation}, most volunteer-driven projects lack this institutional infrastructure. For these smaller or community-led initiatives, high social popularity can quickly become a liability~\cite{pinto2016more}.

Platforms like GitHub allow projects to rapidly accumulate social traction~\cite{dabbish2012social,borges2018githubstars}. Although such visibility is conventionally perceived as a sign of success, it creates a significant social burden for developers and maintainers~\cite{dabbish2012social,dias2021makes}. Because core maintainers operate with strictly finite cognitive bandwidth, the disproportionate influx of casual contributors that accompanies high visibility can be overwhelming~\cite{he2026six}. Research confirms that casual contributors bring tangible net benefits to a project, such as useful bug fixes~\cite{pinto2016more}. However, the sheer volume of general user interactions, such as demanding support or proposing out-of-scope features, can create an ``induced demand" on maintainers~\cite{eghbal2020working}. Such activities inevitably lead to information overload, which drains limited maintainer resources and accelerates burnout~\cite{sutanto2014uncovering}.

This socio-technical tension suggests that project survival should be studied not merely as a function of team size, but as a function of structural accessibility. We define  structural accessibility as the degree to which a project's codebase and documentation lower the cognitive barrier to entry for newcomers. Conventionally, software engineering principles dictate that clean, readable code and external documentation are beneficial for onboarding newcomers~\cite{hunter2021ten,imran2025impact}. Yet, in highly popular projects, we hypothesize that these structural gateways create a ``paradox of accessibility." By effectively lowering entry barriers, highly readable code and extensive wikis act as a magnet for external demands, which amplifies the maintainer's burden. Therefore, to understand modern OSS sustainability, it is essential to investigate how external social demands (measured through social popularity alongside code readability and external wikis) statistically extend or shorten a project’s active development lifespan.

To investigate this paradox of accessibility, we analyzed a stratified random sample of 73,195 collaborative GitHub repositories. We modeled this survival analysis using a robust Accelerated Failure Time (AFT) framework, which effectively neutralizes the right-censoring and time-accumulation biases inherent in open-source lifespan data. To the best of our knowledge, this is the first large‑scale study to empirically decouple open‑source popularity from human labor within a survival framework. While social signals such as stars and forks are often viewed as beneficial indicators~\cite{borges2016understanding,Aggarwal2014}, some studies warn they can also be noisy proxies~\cite{kalliamvakou2014promises, koch2024fault} and sources of maintainer burden~\cite{borges2018githubstars,eghbal2020working}. Building on these critiques, our approach is distinct in conceptualizing them as proxies for coordination overhead~\cite{maldeniya2020herding,pinto2016more}. By jointly examining social popularity, code accessibility, and external documentation (collectively termed ``structural accessibility"), we position our study to reveal how structural features intended to attract users may also introduce maintenance challenges when not accompanied by sufficient labor capacity.

Specifically, we evaluate the impact of external social demand by utilizing a Principal Component Analysis (PCA) construct of joint social popularity metrics (stars, watchers, and forks). We then examine how this massive social demand interacts with a project's structural gateways (operationalized through code readability indices) and the presence of external user documentation (wikis). To confidently isolate these true socio-technical factors of survival, our models rigorously account for various project-level confounding factors.

Our study's contributions are summarized as follows:
\begin{itemize} 
	\item We provide the first large-scale survival analysis of over 73,000 GitHub repositories using an Accelerated Failure Time (AFT) framework, which rigorously accounts for non-proportional hazards and right-censoring in OSS lifespan data. 
	\item We demonstrate that human capital (contributors) is the strongest protective factor for project survival. Conversely, excessive social popularity and the presence of structural gateways such as wikis act as statistical liabilities, suggesting a potential paradox of accessibility. \item We show that interaction effects matter. When the number of contributors interacts with social popularity, the protective effect of labor becomes visible. This highlights the need for governance strategies that balance visibility with labor capacity to ensure long-term OSS resilience. 
\end{itemize}

The rest of the paper is organized as follows. Section~\ref{sec:LR} reviews prior work on hyper‑visibility, structural gateways, and survival factors. Section~\ref{sec:RD} details our research design, including research questions, methodologies, and operationalizations. Section~\ref{sec:Results} presents the results for each research question. Section~\ref{sec:Discuss} synthesizes and interprets these findings in relation to existing knowledge. Section~\ref{sec:Threats} acknowledges limitations and potential threats. Finally, Section~\ref{sec:Concl} concludes the study and outlines directions for future research.

\section{Literature Review}\label{sec:LR}

This section reviews the existing literature regarding the socio-technical drivers of project longevity. We first examine the burden of hyper-visibility and the impact of casual contributors on maintainer resources. We then discuss structural gateways by focusing on how code readability and external documentation facilitate or complicate project accessibility. Finally, we situate this study within the broader context of OSS survival and lifespan modeling.
\subsection{The Burden of Hyper-Visibility and Casual Contributors}\label{subsec:LRHyper}

Historically, the sustainability of OSS projects has relied on the presence of a dedicated core team. A foundational study by Mockus et al.~\cite{mockus2002two} highlighted that successful projects rely on a small, active group of core developers who manage the majority of new functionality. However, the advent of modern social coding platforms like GitHub has fundamentally changed this dynamic by introducing new levels of transparency and hyper-visibility~\cite{dabbish2012social}.

Studies highlight that this hyper-visibility changes contributor behavior. Dabbish et al.~\cite{dabbish2012social} found that visible activity cues allow developers to quickly evaluate project quality. This places the maintainers ``on stage" and draws significant attention towards trending repositories. Consequently, highly popular projects experience a sudden increase in casual contributions~\cite{fatima2025developer}.
Pinto et al.~\cite{pinto2016more} empirically demonstrated that casual contributors can account for 48.98\% of a project's contributor base. While this surge promotes diversity, it also introduces significant coordination overhead. Project maintainers highlight this overhead as a primary source of burden, particularly in the form of out-of-scope features~\cite{dias2021makes,eghbal2020working}. Because maintainer attention is a strictly limited and depletable resource~\cite{avelino2019abandonment}, excessive external demand can lead to information overload and maintainer burnout~\cite{linaaker2024sustaining}. Such factors ultimately threaten project sustainability~\cite{fatima2025developer}. This study builds upon this literature by formally testing whether massive social popularity acts as a liability when controlling for a project's core human capital.

\subsection{Structural Gateways: Readability and External Documentation}\label{subsec:LRRD}

The growing number of casual contributors is largely mediated by a project's structural accessibility~\cite{pinto2016more,linaaker2024sustaining}. Because OSS development relies heavily on a volunteer workforce, developers frequently join, abandon, or casually contribute to projects~\cite{linaaker2024sustaining}. Therefore, code readability becomes a fundamental prerequisite for survival. Recent longitudinal modeling by Tulili et al.~\cite{tulili2025exploring} demonstrates that high contributor turnover poses a severe risk to project continuity and knowledge retention. To mitigate such risks, Fatima et al.~\cite{fatima2025developer} highlight that the source code itself serves as the primary medium of knowledge transfer in open-source environments. A well-structured, well-commented codebase onboards new contributors, whereas convoluted code acts as a severe knowledge barrier~\cite{fatima2025developer}. This aligns with foundational work by Steinmacher et al.~\cite{steinmacher2015social}, who identified code complexity and inadequate documentation as primary technical barriers. Clean, readable code effectively mitigates these barriers, reducing cognitive load and accelerating the onboarding process~\cite{steinmacher2014barriers}. However, under the condition of hyper-visibility, we hypothesize that this frictionless entry creates a `paradox of accessibility.' By making a codebase easier to understand and modify, highly readable code acts as a magnet for casual contributions. This, in turn, increases the review and coordination burden on the core team.

To evaluate this structural accessibility across diverse ecosystems, we utilize language-agnostic structural proxies. Specifically, we use comment density and blank line ratio. As noted by Buse and Weimer~\cite{buse2010readability}, we acknowledge that comments and blank lines are only moderately correlated with human-annotated notions of readability, and thus do not capture logical complexity in its entirety. However, due to the necessity of comparing projects across diverse programming languages where advanced static analysis tools are not universally applicable, these metrics provide reliable, easily operationalizable proxies for typographical or visual readability.

Similarly, the presence of external documentation, such as project wikis, serves as a critical structural gateway. Comprehensive documentation bridges the gap between core maintainers and the broader user base. Dias et al.~\cite{dias2021makes} highlight that up-to-date documentation is essential, as it allows newcomers to easily understand project requirements and infrastructure that the core team might otherwise take for granted. However, this high accessibility also transforms documentation into a primary target for `drive-by' commits. Pinto et al.~\cite{pinto2016more} empirically demonstrated this. They found that a significant portion of all casual contributions (nearly 29\%) are directed specifically at fixing, translating, or updating documentation. While wikis and documentation theoretically facilitate initial onboarding, managing these incoming contributions and ensuring continuous maintenance require sustained effort from the core team. Therefore, in this study, we investigate the Documentation Paradox. We hypothesize that while structural gateways like wikis are intended to reduce support overhead, they act as magnets for casual contributor demands, which then amplifies the maintenance liability in highly popular projects.

\subsection{OSS Survival and Lifespan Modeling}\label{subsec:LRSurvival}

The long-term viability of OSS projects is traditionally measured through survival analysis. Foundational work by Samoladas et al.~\cite{Samoladas2010} applied Cox Proportional Hazards models to demonstrate that the number of active committers is a highly significant predictor of project survival. Subsequent studies expanded this to include dynamic predictors. For instance, Wang et al.~\cite{Wang2012} demonstrated that the importance of survival predictors changes as projects mature, highlighting the limitations of static models. More recently, Park et al.~\cite{park2024assessing} utilized Kaplan-Meier estimators and polynomial regression to empirically prove that community engagement and social popularity metrics (e.g., stars and forks) have a strong, statistically significant correlation with project survivability.

Furthermore, our own previous work~\cite{Kaushik2025} explored the multi-dimensional nature of this popularity using Exploratory Factor Analysis. That study highlighted that community engagement must be deconstructed into active and passive components. The current study directly extends this into an AFT survival framework. In doing so, we draw on the findings of Joblin and Apel~\cite{Joblin2022}, who suggested that framing project outcomes through a `joint socio-technical' lens is significantly superior to traditional developer-centric models. Despite this, the socio-technical interplay remains an under-explored gap in lifespan modeling. To our knowledge, no prior study has formally integrated external social demand with internal measures of structural accessibility, such as technical gateways like code readability and wikis, to determine an OSS project's active maintenance lifespan.

\section{Research Design Overview}\label{sec:RD}

This section outlines our research design, which is guided by our primary research questions. It provides the necessary details regarding data collection, preprocessing, and the operationalization of our variables. Finally, we describe the specific methodologies used to evaluate the socio-technical drivers of project survival.

\subsection{Research Questions}\label{subsec:RQs}

To address the socio-technical gap in OSS lifespan modeling and to ensure the robustness of our findings, we formulated the following research questions:
\begin{itemize}
	\item RQ1 (Main Effects): How do social popularity, code readability, and external documentation affect the longevity of OSS projects?
	
	\textbf{Methodology:} Survival modeling frameworks were used to evaluate the independent impact of each socio-technical construct on project duration. This parametric approach allows for the estimation of the `acceleration' or `deceleration' of a project's expected lifespan in response to specific covariates.
	
	\item RQ2 (Moderation Effects): To what extent do human capital, code readability, and the presence of documentation moderate the relationship between social popularity and project longevity?
	
	\textbf{Methodology:} We extended the existing survival model by incorporating three primary interaction terms: 
	\textit{Social Popularity $\times$ Labor}, \textit{Social Popularity $\times$ Readability}, and \textit{Social Popularity $\times$ Wiki}. 
	These terms allow us to test whether internal structural and human resources amplify or diminish the impact of social popularity. Specifically, the \textit{Social Popularity $\times$ Labor} interaction examine whether human capital can buffer social demand, while the readability and wiki interactions evaluate the paradox of accessibility. By analyzing these coefficients, we determine if high structural accessibility acts as a protective force multiplier for project longevity or if it creates a maintenance liability under conditions of hyper‑visibility.
	
	\item RQ3 (Sensitivity Analysis): How sensitive are these observed survival effects to varying operational definitions of project inactivity?
	
	\textbf{Methodology:} To ensure the temporal stability of our predictors, we performed a robustness check by re-estimating the survival models across four distinct inactivity thresholds: 3, 6, 9, and 12 months. This approach allows us to verify that our findings are not sensitive to a specific definition of project inactivity. By comparing the consistency of the coefficients across these varying temporal windows, we can confirm the reliability of our socio-technical constructs.
\end{itemize}

\subsection{Data Collection and Filtering}\label{subsec:DC}

To initiate our analysis, we retrieved the initial dataset from the SEART GitHub Search (GHS) platform~\cite{Dabic2021}.\footnote{\url{https://seart-ghs.si.usi.ch/}} This comprehensive dataset contains information on 1.77 million repositories, with each entry characterized by 35 distinct attributes. These attributes include both technical metadata, such as primary programming language and file counts, and social metrics, such as star and fork counts.

Because the GHS platform indexes public GitHub repositories, it contains many experimental projects, forks, and projects by solo developers. To isolate sustainable, community-driven software and ensure the validity of our survival analysis, we formulated the following exclusion criteria:

\begin{itemize}
	\item We excluded all repository mirrors and forks to ensure that each project represents an independent, original development effort rather than a derivative of existing work. This criterion is necessary because forks often possess different streams of development and project lifespans~\cite{jiang2017}. In this step, we identified 58,149 fork repositories.
	\item Collaborative Development: To ensure that projects are collaboratively developed, we excluded repositories with fewer than three contributors or zero pull requests~\cite{kalliamvakou2014promises,xia2022predicting}. We identified 637,435 repositories failing to meet this criterion.
	\item Baseline Development: To ensure that repositories have a baseline of activity, we filtered out those with fewer than 10 commits~\cite{kalliamvakou2014promises,robles2025comparative}. We identified 332,460 repositories lacking this baseline activity.
	\item Formal Releases: We focused on repositories that have at least one formal release. This criterion was developed for two reasons. First, after an initial release, projects typically attract a larger number of contributions~\cite{Midha2012}. This directly aligns with our objective of testing the effect of hyper-visibility. Second, survival factors in early phases of development differ from those in long-term survival~\cite{Midha2012,zhou2012make}. We identified 545,211 repositories lacking a release.
	\item No License: Licensing plays a crucial role in project popularity, and we consider it a control variable~\cite{Sen2012}. We filtered out projects that lack a valid license. We identified 570,978 repositories with invalid licenses.
	\item Language Standardization: To ensure metric compatibility, ecosystem coverage, and statistical power, we restricted our sample to the top programming languages. While our raw structural metrics are language-agnostic in their extraction, they must be standardized within language cohorts to ensure metric compatibility across the dataset (as detailed in Section~\ref{subsubsec:IVs}). Restricting our sample to the most prominent languages ensures we have sufficiently large and representative cohorts to perform this within-language normalization. These languages dominate GitHub in terms of repository count and user engagement. Prior studies demonstrate that a small subset of languages accounts for the majority of lines of code, project appearances, and developer activity~\cite{bissyande2013popularity}. Hoffmann et al.~\cite{hoffmann2024value} estimate that the top six languages, as presented in Table~\ref{tab:language_rankings} contribute 84\% of the global demand-side value of OSS. Thus, this filtering step allows us to preserve comprehensive ecosystem coverage while supporting statistically sound metric standardization.
	\end{itemize}
	
	Languages were selected based on their dual prominence in Hoffmann et al.'s OSS value rankings and RedMonk's 2025 developer engagement index~\cite{redmonk2025rankings}. The Table~\ref{tab:language_rankings} summarizes this selection:
	
	\begin{table}[!h]
		\caption{Top 10 Programming Languages by OSS Value (Hoffmann et al.) and Developer Engagement (RedMonk)}
		\label{tab:language_rankings}
		\centering
		\begin{tabular}{lcc}
			\toprule
			\textbf{Language} & \textbf{Hoffmann (2024)} & \textbf{RedMonk (2025)} \\
			\midrule
			JavaScript & \checkmark & 1 \\
			Python     & \checkmark & 2 \\
			Java       & \checkmark & 3 \\
			PHP        & \checkmark & 4 \\
			C\#        & \checkmark & 5 \\
			TypeScript & \checkmark & 6 \\
			C++        & ---        & 7 \\
			Ruby       & ---        & 8 \\
			C          & ---        & 9 \\
			Go         & ---        & 12 \\
			\bottomrule
		\end{tabular}
	\end{table}
	
	Languages structurally incompatible with our modeling pipeline, such as CSS and Swift, were excluded. CSS is a declarative language lacking procedural logic and is not amenable to readability metrics like comment density or blank line ratio~\cite{bosch2014css}. Swift was excluded due to its tight binding to closed-source Apple ecosystems and frequent compiler instability in early repositories~\cite{reboucas2016swift,sonarsource2021swift}. Such factors complicate metric extraction and survival modeling.
	
	To ensure comparability across ecosystems, we applied stratified random sampling by primary programming language. Within each language stratum, repositories were randomly selected and balanced by project activity state. Specifically, we drew an equal number of active and inactive repositories. After applying all criteria, stratifying by language, and balancing by activity, our final sample consists of 73,195 repositories across 9 programming languages. Ruby was excluded during the final sampling phase due to insufficient representation of active projects after all filters were applied. Ruby's insufficient representation in our sample can be linked to its Gem-centric workflow. As identified by Decan et al.~\cite{decan2019empirical}, the true lifecycle of releases and dependencies in the Ruby ecosystem is governed by the RubyGems package manager rather than source code repositories. This occurs because Ruby developers frequently treat RubyGems.org as their primary release authority instead of utilizing GitHub's formal release tags. Consequently, a disproportionate number of active Ruby repositories failed to meet our GitHub-based release criteria.
	
	\subsection{Construct Definitions and Operationalization}\label{subsec:Operationalization}
	
	In this section, we provide formal definitions of the constructs and outline the operationalization and mathematical transformations applied to our dependent, independent, and control variables.
	
	\subsubsection{Dependent Variable}\label{DV}
	
	Defining project end of active development in OSS is challenging because repositories rarely use official archival features. Researchers therefore typically use inactivity thresholds as a proxy for termination. However, we acknowledge that an absence of commits conflates true project abandonment with successful project completion (e.g., a stable, feature-complete utility).Furthermore, no consensus exists on a specific duration~\cite{Evangelopoulos2008,Calefato2022}. We define a project ``event'' ($E=1$) as being explicitly archived or having zero commits in the six months preceding data collection~\cite{Liao2019}. Conversely, we treat active projects as right-censored observations ($E=0$). This distinction prevents biases in time accumulation and allows the model to account for ongoing projects. To ensure robustness, we also tested 3, 9, and 12-month thresholds.
	
	\subsubsection{Independent Variables}\label{subsubsec:IVs}
	
	To assess the socio-technical dimensions of OSS survival, we consider two primary constructs: social popularity, and structural accessibility.
	\begin{itemize}
		\item Social Popularity: We utilized GitHub’s native social indicators, specifically stargazers ($S$), forks ($F$), and watchers ($W$). Since the literature identifies a high correlation among these indicators~\cite{borges2016understanding}, we defined Social Popularity as a principal component index derived from these measures:

		\[
		\text{Social Popularity}_{\text{raw}} = w_{1}S + w_{2}F + w_{3}W
		\]

		where $w_{1}, w_{2}, w_{3}$ are the loadings obtained from Principal Component Analysis (PCA).
		
		Because PCA scores may be negative and the direct logarithm of zero is undefined, we treated the construct by shifting the scores to ensure non‑negativity and applying the \texttt{log1p} transformation~\cite{karakaplan2020solution,ohaegbulem2024remedying}:

		\[
		\resizebox{\linewidth}{!}{$
			\text{Social Popularity} = \ln\!\Big(1 + \big(\text{Social Popularity}_{\text{raw}} - \text{PCA}_{\min}\big)\Big)
			$}
		\]

		This treatment ensures that the popularity scores remain strictly non‑negative and reduces the impact of extreme outliers common in OSS data.
		
		\item Structural Gateways: We measure the structural accessibility using two distinct proxies:\\
			1) Readability Index: Code readability was quantified using two structural metrics: comment density and blank line ratio. The underlying attributes, \texttt{codeLines}, \texttt{commentLines}, and \texttt{blankLines}, were provided directly by the SEART platform, which precomputes these values using lightweight parsing techniques that do not require compilation or language-specific tooling. Using these pre-extracted counts, we computed the readability metrics using the following formulas:
			
			\begin{align*}
				\text{Comment Density} &= \frac{\texttt{commentLines}}{\texttt{codeLines}} \\
				\text{Blank Line Ratio} &= \frac{\texttt{blankLines}}{\texttt{codeLines}}
			\end{align*}
			
			To account for language-specific formatting and documentation norms, each metric was standardized within language groups using z-score normalization. The final Readability Index was computed as:
			
			\[
			\resizebox{\linewidth}{!}{$
				\text{Readability Index} = \frac{z(\text{Comment Density}) + z(\text{Blank Line Ratio})}{2}
				$}
			\]
			
			As z-score standardization centers the data without altering its underlying distributional shape, the resulting index naturally retains the severe right-skewness characteristic of the raw code metrics. Furthermore, z-scores inherently introduce negative values. To stabilize the variance for our survival models and resolve the negative scale, we systematically applied a non-negative minimum shift (subtracting the minimum observed value from each score) followed by a logarithmic transformation (log1p). 
			
			We applied this within-language normalization because structural conventions vary significantly across ecosystems. For instance, the commenting patterns in C code are not directly comparable to the verbosity norms of Java or Python~\cite{prechelt2002empirical}. By standardizing these metrics, every repository's code readability is compared to its peers written in the same language. This approach ensures that the index captures project-level accessibility relative to community standards rather than reflecting inherent language-level differences.
			
			Furthermore, we assigned equal weights to these components based on our theoretical understanding that internal documentation and visual structure contribute equally to overall code readability. This decision is supported by prior empirical work which identifies both comment density and blank line ratio as key typographical elements that positively correlate with code comprehensibility \cite{fakhoury2019improving, mohan2004programming}. While some studies suggest varying levels of local importance for these features \cite{buse2010readability}, we treat them as complementary dimensions of a project's structural accessibility, where comments provide semantic explanation and blank lines provide logical chunking of the source code~\cite{mohan2004programming,stegeman2014towards}. While more advanced metrics such as cyclomatic complexity offer deeper insights into control flow and logical branching, they were not included in this study due to tooling and scalability constraints. Extracting such metrics requires language-specific static analysis tools (e.g., SonarQube, Radon, Lizard), which depend on buildable codebases and consistent project structures. These conditions often unmet in large-scale, cross-language OSS datasets.
			
			2) External Documentation: This was captured using the the \texttt{hasWiki} attribute, a binary indicator of whether the repository includes a GitHub Wiki. We focused specifically on the Wiki as an indicator of external support, deliberately excluding internal documentation signals such as README content or inline comments. READMEs are typically concise and developer-focused, providing minimal build instructions and setup notes~\cite{github_readme_docs}. In contrast, a GitHub Wiki is a separate, voluntary feature used for more comprehensive, user-oriented guides, tutorials, and usage scenarios that extend beyond the technical requirements for interacting with the source code. Also, we acknowledge that this binary flag only indicates structural presence and cannot measure the quality, maintenance level, or actual usage of the documentation.
	\end{itemize}
	
	\subsubsection{Control Variables}\label{subsubsec:CV}
	
	To isolate the main and moderating effects of our independent variables, we include several project-level controls known to influence OSS survival baselines:
	
	\begin{itemize}
		\item Project Size: Measured as the total lines of code (log-transformed). This controls for the technical complexity and scale of the project. Larger projects often exhibit different growth patterns and maintenance requirements compared to smaller utilities~\cite{Samoladas2010,Liao2019}.
		\item Total Contributors: We control for the project's human capital by including the log-transformed count of unique contributors. A larger developer base is a primary predictor of survival, as it provides the necessary labor for bug fixes and feature development~\cite{Samoladas2010,Wang2012}.
		\item Release Frequency: Calculated as the average number of formal releases per month (log-transformed), this variable accounts for development momentum. Consistent release cycles are a key indicator of project health and associated with continued active development~\cite{wiggins2010reclassifying,subramaniam2009determinants}.
		\item Programming Language: We include a categorical control for the primary programming language. This accounts for ecosystem-specific survival trajectories and the fact that certain languages (e.g., C or Java) may attract different levels of developer interest and user demand \cite{Sen2012,sen2015application}.
		\item License Category: Projects are categorized by license restrictiveness (e.g., Permissive vs. Copyleft). The choice of license dictates legal adoption barriers and has been shown to significantly influence a project's ability to attract long-term contributors \cite{Midha2012, Sen2012}.
		\item Project Creation Year: This categorical control accounts for ``cohort effects.'' It adjusts for varying baseline ages, changes in the maturity of the GitHub ecosystem, and historical shifts in platform competition that might affect the survival probability of projects founded in different eras \cite{wiggins2010reclassifying,schweik2012internet}.
	\end{itemize}
	
	Because social signals accumulate over time, we include Project Creation Year as a cohort control to mitigate age-related variance. However, we deliberately avoid normalizing these metrics by project lifespan. Since lifespan is the explicit dependent variable in our survival models, including it in the denominator of our independent predictors would introduce severe endogeneity\footnote{In regression and survival models, endogeneity occurs when a predictor variable is correlated with the error term, often due to mathematical coupling.}. This mathematical coupling creates a spurious inverse relationship that effectively penalizes long-lived projects. To rigorously account for time-accumulation without violating survival modeling assumptions, we retain cumulative counts and rely on the fixed creation year to control for exposure time and ecosystem maturity.
	
	\subsection{Analytical Framework}\label{subsec:AFT}
	
	To model the active maintenance lifespan of OSS projects and accommodate right-censored data, we employ survival analysis. While the Cox Proportional Hazards (PH) model is a common semi-parametric approach, it relies on the foundational assumption that the hazard ratio between subjects remains constant over time~\cite{moran2008modelling}. However, socio-technical factors like community engagement and structural complexity often exhibit time-varying effects as projects mature~\cite{sutanto2014uncovering,qi2009comparison}. Therefore, we first formally test the proportional hazards assumption using Schoenfeld residuals~\cite{orbe2002comparing}.
	
	To ensure methodological robustness in the event of PH violations, we utilize a parametric AFT framework. Also, it is helpful to contrast AFT with the traditional Cox model. While Cox models estimate the hazard rate (the immediate risk of failure), AFT models do not assume proportional hazards. Instead, they directly evaluate the natural logarithm of the survival time itself, which makes them highly effective for capturing dynamic software lifespans\cite{moran2008modelling,qi2009comparison}. Conceptually, the AFT framework assumes that the effect of an independent variable is to multiply the expected lifespan by a constant factor, effectively ``accelerating" or ``decelerating" the passage of time~\cite{orbe2002comparing,louis1981nonparametric}.
	
	Mathematically, this relationship is expressed by modeling the natural logarithm of the survival time \(T\) as a linear function of the covariates \(X\):

	\[
	{\ln(T) = \beta_{0} + \beta_{1}X_{1} + \beta_{2}X_{2} + \dots + \beta_{n}X_{n} + \sigma \varepsilon}
	\]

	In this equation, \(\beta\) represents the coefficients for the given variables, \(\sigma\) is the scale parameter, and \(\varepsilon\) is the error term whose specific distribution defines the survival model.
	
	Furthermore, exponentiating these coefficients \(\left(e^{\beta}\right)\) yields \emph{Time Ratios (TR)}. These are highly interpretable metrics that indicate how specific covariates accelerate or decelerate the time to project inactivity \cite{louis1981nonparametric, patel2006comparing}. A \(TR > 1\) implies that a covariate stretches or prolongs a project’s active lifespan (a protective effect), whereas a \(TR < 1\) implies that time is compressed, accelerating the project's cessation \cite{patel2006comparing}.
	
	We fit and compare standard survival distributions, including Weibull, Log-Normal, and Log-Logistic. The final model is selected based on the Akaike Information Criterion (AIC) and log-likelihood goodness-of-fit scores.

	\section{Results}\label{sec:Results}
	This section presents our findings by research question. First, we describe the dataset and construct the indices for latent social popularity and code readability. We then detail the selection and diagnostics of the survival model before answering the research questions.
		
	\subsection{Descriptive Statistics}\label{DS}
	After applying all inclusion criteria, language stratification, and activity balancing, our final sample consists of 73,195 collaborative OSS repositories. Table~\ref{tab:descriptivestats} presents the descriptive statistics for the continuous variables.
	
	\begin{table*}[!h]
		\centering
		\caption{Descriptive Statistics for Continuous Variables}
		\label{tab:descriptivestats}
		\begin{tabular}{lrrrrrr}
			\toprule
			Variable & Mean & Median & Std.\ Dev. & Min & Max & Skewness \\
			\midrule
			\multicolumn{7}{l}{\textbf{Dependent Variable}} \\
			Lifespan (months) & 63.82 & 57.49 & 40.73 & 0.01 & 213.90 & 0.64 \\
			\midrule
			\multicolumn{7}{l}{\textbf{Raw Social Metrics}} \\
			Stargazers & 681.63 & 75.00 & 3,329.07 & 10.00 & 191,982.00 & 18.62 \\
			Forks & 121.67 & 20.00 & 849.56 & 1.00 & 74,908.00 & 46.20 \\
			Watchers & 24.85 & 9.00 & 84.43 & 1.00 & 7,445.00 & 25.34 \\
			\midrule
			\multicolumn{7}{l}{\textbf{Raw Readability Metrics}} \\
			Comment Density & 0.18 & 0.12 & 0.94 & 0.00 & 181.88 & 129.57 \\
			Blank Line Ratio & 0.18 & 0.17 & 0.27 & 0.00 & 57.91 & 151.27 \\
			Readability Index (z-score avg) & 0.00 & -0.05 & 0.83 & -1.76 & 76.71 & 18.72 \\
			\midrule
			\multicolumn{7}{l}{\textbf{Control Variables}} \\
			Total Contributors & 22.52 & 8.00 & 113.20 & 3.00 & 9,151.00 & 55.82 \\
			Releases per Month & 0.88 & 0.24 & 4.48 & 0.00 & 335.82 & 35.43 \\
			Repository Size (KB) & 48,612.39 & 3,187.00 & 276,807.90 & 6.00 & 22,846,100.00 & 31.09 \\
			\bottomrule
		\end{tabular}
	\end{table*}
	
	From Table~\ref{tab:descriptivestats}, the dependent variable, Lifespan, exhibits a mean of 63.82 months and a median of 57.49 months (approximately 4.8 years). The maximum value reaches nearly 18 years, indicating the presence of highly mature projects within the sample. The skewness (0.64) and the similarity between the mean and median show that the data distribution is balanced. This balance confirms that the target variable is not disproportionately skewed by either short-lived experimental projects or extremely mature ``mega-projects.'' Unlike lifespan, the raw social metrics highlight extreme skewness. While the median project receives modest engagement, the mean values are much higher by a fraction of outlier projects. The large difference between the median and mean shows severe right-skewness for stargazers, forks, and watchers. This confirms that social visibility is concentrated among a small number of repositories. These distributions require PCA and logarithmic scaling to ensure the data meet the assumptions of the survival model.
	
	Structural code attributes show different scales and standard deviations for comment density and blank line ratio. Both show severe right-skewness because some repositories contain mainly auto-generated comments or aggressive formatting of whitespace. This disparity requires z-score standardization for each language, as languages follow different structural rules. As expected, index normalization reduces this skewness by setting the mean to zero. However, extreme outliers remain after z-scoring, which confirms the need for further transformations.
	
	Development intensity and project scope are similarly concentrated. Most projects have few contributors and small repositories. However, a few large codebases increase the maximum values. This imbalance produces high skewness, showing that raw inputs of development scale multiplicatively. This behavior justifies applying logarithmic transformations to all continuous variables for control. By stabilizing the variance, we ensure that a few massive projects do not bias the survival model.
	
	We then analyze the categorical statistics of our sample. Table~\ref{tab:frequencydistribution} summarizes these frequency distributions across documentation, license, and language categories, stratified by project state.
	
	\begin{table}[!h]
		\centering
		\caption{Frequency Distribution for Categorical Variables by Project State}
		\label{tab:frequencydistribution}
		\begin{tabular}{lrr}
			\toprule
			\textbf{Variable / Category} & \textbf{Active (\%)} & \textbf{Inactive (\%)} \\
			\midrule
			\multicolumn{3}{l}{\textbf{Documentation}} \\
			Has Wiki (True)  & 64.59 & 75.43 \\
			No Wiki (False)  & 35.41 & 24.57 \\
			\midrule
			\multicolumn{3}{l}{\textbf{License Category}} \\
			Permissive & 65.90 & 69.12 \\
			Copyleft   & 20.76 & 17.97 \\
			Other      & 13.34 & 12.91 \\
			\midrule
			\multicolumn{3}{l}{\textbf{Primary Language}} \\
			Go          & 12.40 & 12.60 \\
			TypeScript  & 12.02 & 12.09 \\
			C\#         & 11.59 & 12.06 \\
			C++         & 12.08 & 11.49 \\
			PHP         & 10.70 & 12.48 \\
			Python      & 11.93 & 11.07 \\
			C           & 11.97 & 10.95 \\
			Java        & 11.27 & 10.16 \\
			JavaScript  & 6.02  & 7.12 \\
			\bottomrule
		\end{tabular}
	\end{table}
	
	Table~\ref{tab:frequencydistribution} reveals some interesting patterns. The distribution of project documentation reveals a counter-intuitive pattern. While 64.59\% of active projects maintain a wiki, this figure rises to 75.43\% for inactive projects. These percentages suggest that wiki presence is not a definitive predictor of ongoing activity. Instead, this trend initially aligns with our paradox of accessibility hypothesis. The higher prevalence of wikis in inactive projects suggests that the presence of such documentation might act as a structural liability rather than a protective factor. This dynamic is formally tested in our subsequent survival models.
	
	Regarding licensing, permissive licenses are the most common across both states, representing over 65\% of the sample. The proportion of copyleft licenses is slightly higher in active projects. This suggests that while permissive licenses dominate the ecosystem, projects with copyleft obligations may show different survival characteristics. Finally, the distribution of primary languages is consistent across both active and inactive states. Most languages maintain a representation between 10\% and 12\%. This indicates that our stratification strategy successfully balanced the language ecosystems. This balance ensures that our survival models are not biased toward a specific programming community.
	
\begin{figure}[!h]
		\centering
		\includegraphics[width=\linewidth]{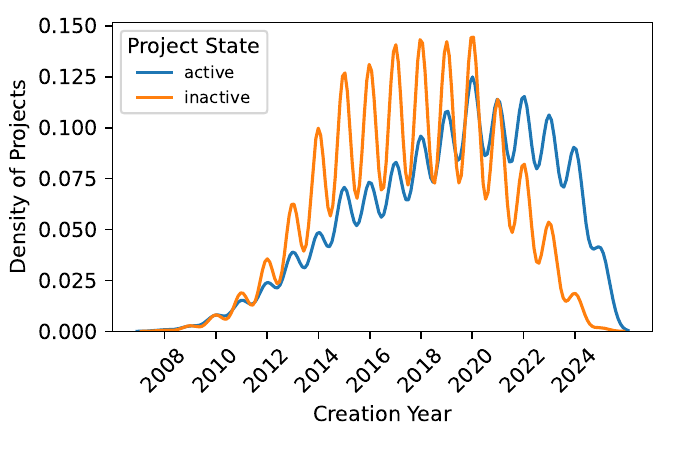}
		\caption{Kernel Density Estimation of Project Creation Year by State}
		\label{fig:kdeprojectstate}
	\end{figure}

	Temporal bias is a common threat to observational survival studies. This bias arises from the assumption that the active cohort consists of recent projects while the inactive cohort contains older repositories~\cite{kalliamvakou2014promises,Samoladas2010}. We test this assumption by visualizing the temporal dynamics of the dataset. Figure~\ref{fig:kdeprojectstate} shows the Kernel Density Estimation (KDE) of project creation years, stratified by activity state.
	
	The distribution shows that our analysis does not merely compare dormant projects from 2008 against active projects from 2024. Instead, active and inactive repositories coexist across the 17-year timeline. The diverging density curves reveal a demographic shift in the ecosystem of open-source software. Older cohorts maintain a balanced ratio of active to inactive states. However, the density of inactive projects spikes between 2014 and 2020 and overtakes active projects. We acknowledge a limitation in the most recent cohorts (2020+). The apparent rightward shift toward ``active" projects is affected by length-biased sampling. Currently, these recent cohorts only capture fast-dying projects. Slow-dying projects simply have not had sufficient time to transition to inactivity and be observed. Thus, this recent divergence is partly an artifact of the observation window. Despite this limitation, the overall coexistence of both states across the broader 17-year timeline effectively mitigates general right-censoring concerns. Overall, these results refute the temporal bias assumption.
	
	\subsection{Latent Construct Extraction: Social Popularity (PCA)}\label{subsec:PCA}
	
	We operationalize Social Popularity using a PCA construct. Prior to extraction, we tested for multicollinearity by applying the Variance Inflation Factor (VIF). The VIF values ranged from 2.33 to 2.41, which indicate moderate collinearity. We then applied the Kaiser--Meyer--Olkin (KMO) test, which revealed a value of $0.743$. Bartlett’s test of sphericity also showed high significance, with a chi-square value of $\chi^{2} = 113{,}462.64$ ($p < 0.001$). The PCA successfully extracted a single dominant latent component with an eigenvalue of $2.41$. This single factor captures $80.19\%$ of the total variance in the social metrics. The component loadings were balanced: stargazers ($0.579$), watchers ($0.578$), and forks ($0.575$). These results confirm that these three visibility metrics do not act independently. Instead, they form a singular, unified dimension of external social popularity. This high level of explained variance supports our decision to collapse these metrics into a single index, reducing the complexity of the final survival model. Furthermore, descriptive statistics for the raw PCA index revealed a minimum value of -0.362 and a skewness of 26.13. Such a high level of skewness confirms that even after extracting the latent dimension, the distribution remains heavily concentrated among a few highly visible projects. We addressed this skewness by applying the shifting and log transformation detailed in Section~\ref{subsubsec:IVs}. This final refinement stabilizes the variance of the index, ensuring its suitability for the subsequent survival analysis.
	
	\subsection{Non-Parametric Survival Estimates (Kaplan-Meier)}\label{subsec:KMAnalysis}
	
	We used the Kaplan-Meier (KM) estimator as a baseline before multivariate modeling. To generate survival estimates, we stratified the sample across various socio-technical categories, including social popularity and structural documentation. This non-parametric test determines if the survival probabilities between strata differ significantly.
	
\begin{figure}[!h]
		\centering
		\includegraphics[width=\linewidth]{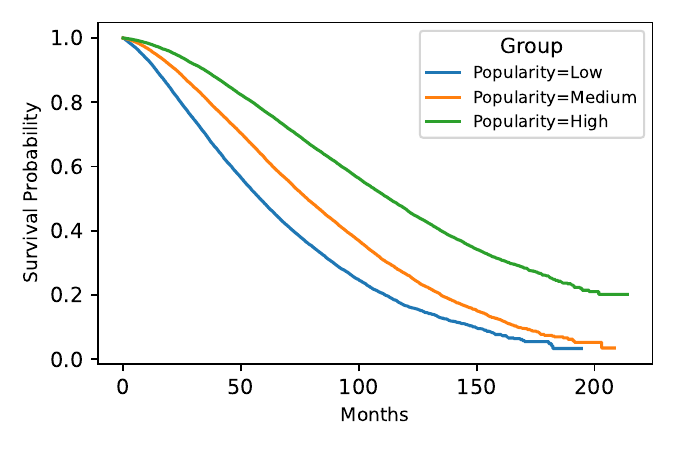}
		\caption{Kaplan-Meier survival estimates stratified by Social Popularity tertiles.}
		\label{fig:kmsocialpopularity}
	\end{figure}

	\paragraph*{Social Popularity and Structural Attributes:} To evaluate continuous variables in this non-parametric context, we divided Social Popularity and Readability into tertiles (Low, Medium, High). Additionally, we computed the median survival time, representing the point at which half of the population has experienced the the cessation of active development. Figure~\ref{fig:kmsocialpopularity} visualizes the social popularity tertile curves.
	
\begin{figure}[!h]
		\centering
		\includegraphics[width=\linewidth]{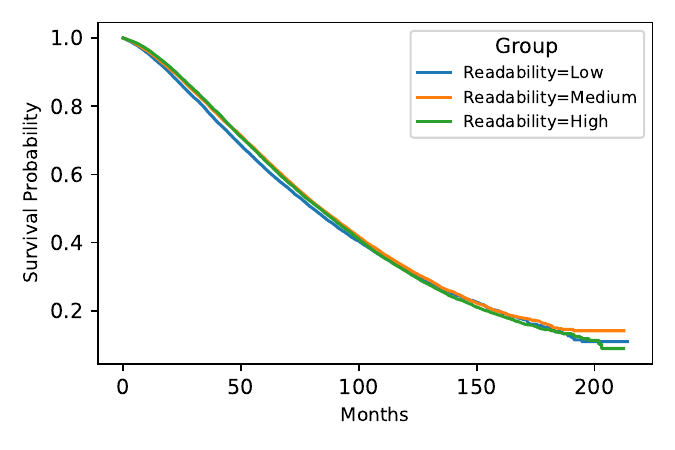}
		\caption{Kaplan-Meier survival estimates stratified by Code Readability tertiles.}
		\label{fig:kmreadability}
	\end{figure}

	From Figure~\ref{fig:kmsocialpopularity}, we observe a clear separation of survival probabilities across the popularity tertiles. The calculated median survival times are as follows: the High Popularity tertile reached 112.90 months, nearly double the median lifespan of Low Popularity projects (58.04 months). These results align with the visual patterns observed in the survival curves. At this exploratory stage, the findings highlight social popularity as a critical lifeline for project longevity. Conversely, we observed an overlapping of the survival curves for the Code Readability tertiles (Figure~\ref{fig:kmreadability}). Median survival times for code readability are 80.57, 83.99, and 83.41 months for the low, medium, and high tertiles, respectively. The log-rank test is not significant ($p = 0.11$), showing no difference between the medium and high tertiles. Thus, code readability alone may not drive project longevity.
	
\begin{figure}[!h]
		\centering
		\includegraphics[width=\linewidth]{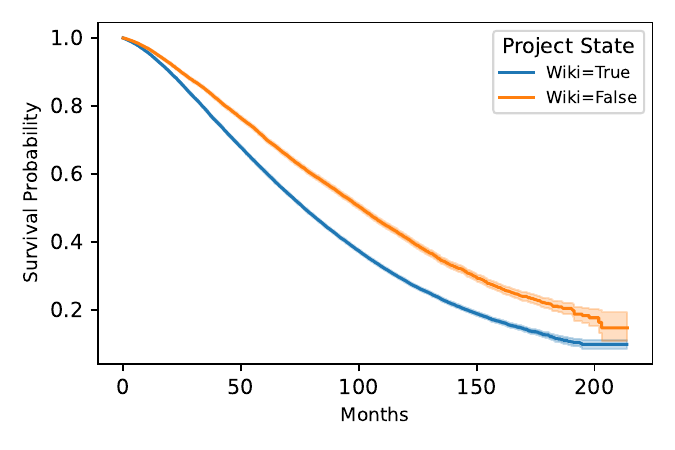}
		\caption{Kaplan-Meier survival estimates stratified by Wiki presence.}
		\label{fig:kmwikisurvival}
	\end{figure}

	Additionally, we modeled the survival curves for Wiki Presence. Figure~\ref{fig:kmwikisurvival} shows a clear separation between the groups, confirmed by a significant p-value from the log-rank test. 
	
	Projects with a wiki show a shorter median survival time (76.67 months) than those without (100.95 months). We initially attributed this difference to recent wiki adoption. However, the distribution across creation years refutes this. By 2010, the dataset already contained 136 active and 165 inactive projects with wikis. By 2015, these counts increased to 1,300 active and 2,800 inactive projects. These data show that wikis are not a recent phenomenon. This finding aligns with our initial observations of wiki prevalence in inactive projects.
	
\begin{figure}[!h]
		\centering
		\includegraphics[width=\linewidth]{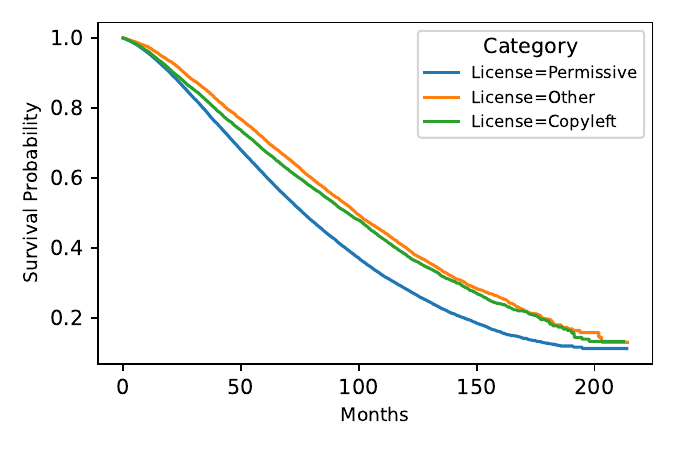}
		\caption{Kaplan-Meier survival estimates stratified by License Category.}
		\label{fig:kmlicensecategory}
	\end{figure}

\begin{figure}[!h]
		\centering
		\includegraphics[width=\linewidth]{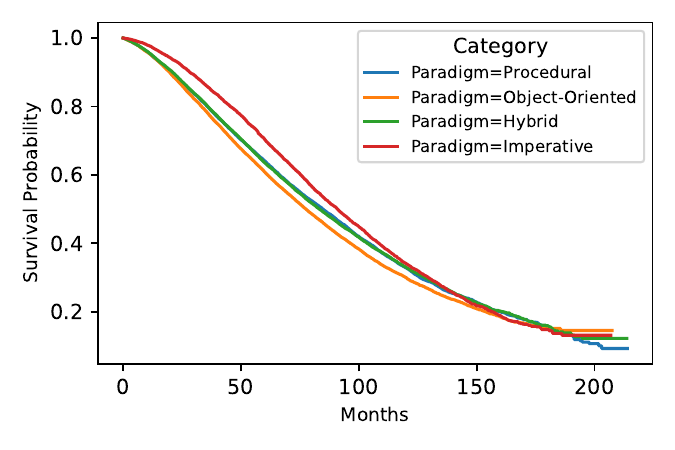}
		\caption{Kaplan-Meier survival estimates stratified by Programming Paradigm.}
		\label{fig:kmprogrammingparadigm}
	\end{figure}

	\paragraph*{Governance and Ecosystem Attributes:} We also visualize these KM curves across project license categories and programming paradigms. Figures~\ref{fig:kmlicensecategory} and \ref{fig:kmprogrammingparadigm} demonstrate a mixed relationship among these categorical attributes. For licensing, we observe that projects with Permissive licenses exhibit a median survival of 76.5 months, which is lower than Copyleft (95.36 months) and Other category licenses (98.93 months).
	
	Regarding programming paradigms, the results show varying levels of resilience. Projects utilizing the Imperative paradigm show the highest median survival at 90.97 months, followed by Procedural (84.8 months) and Hybrid (83.19 months) paradigms. Interestingly, Object-Oriented projects exhibit the lowest median survival in this cohort at 77.59 months. Log-rank tests confirm that the differences in survival trajectories are significant. These results suggest that a project's technical and legal frameworks affect its longevity.
	
	Overall, these exploratory results initially appear to challenge our hypothesis regarding hyper-visibility by framing social popularity as a vital lifeline. However, the Kaplan-Meier estimator is a univariate approach with inherent limitations~\cite{kaplan1958nonparametric}. It cannot account for the simultaneous impact of multiple covariates, such as project age (temporal confounding) and human labor capacity. Since KM curves do not isolate confounding effects, these exploratory findings alone cannot reject the primary hypothesis. We therefore use a multivariate framework to separate popularity from structural factors.
	 
	\subsection{Model Diagnostics and Selection}\label{subsec:ModelSelection}
	
\begin{figure*}[!h]
	\centering
	\includegraphics[width=0.8\linewidth]{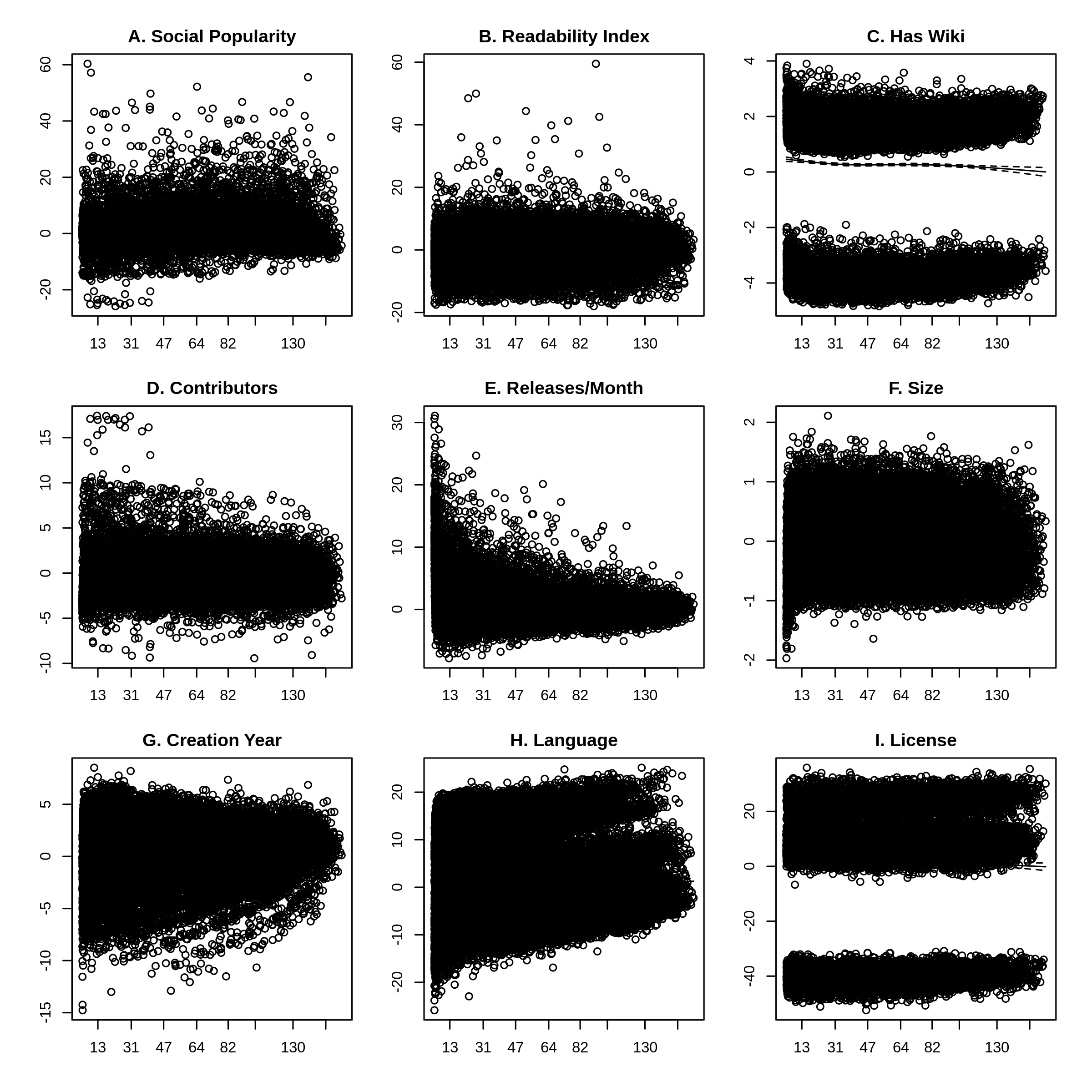}
	\caption{Schoenfeld residual diagnostic plots for the Cox proportional hazards model covariates.}
	\label{fig:phdiagnosticsgrid}
\end{figure*}

	To model project lifespan, we initially fitted the Cox Proportional Hazards (PH) model. 
	We evaluated the constant hazard assumption using the Schoenfeld residual diagnostic plots presented in Figure~\ref{fig:phdiagnosticsgrid}. Each panel in the grid displays the estimated time-varying coefficient $\beta(t)$ for a specific predictor. The results show violations across our core predictors. These appear as deviations from the horizontal zero-line. Specifically, the test for non-proportionality yielded $p < 0.005$ for human capital, structural documentation, development momentum, and our standardized code metric. The global test confirms these failures ($\chi^2 = 26245.87$, $df = 33$, $p < 2\times10^{-16}$). This result violates the proportional hazards assumption. Thus, hazard rates in our dataset vary over time.
	This aligns with theoretical expectations of software ecosystems, where the risk of project inactivity fluctuates as platforms saturate and technologies age. Consequently, relying on Cox PH estimates would yield invalid and biased coefficients. These findings necessitated a transition to a parametric approach, which does not require the proportional hazards assumption.

	To ensure methodological robustness, we transitioned to AFT models. Specifically, we evaluated the Weibull, Log-Normal, and Log-Logistic distributions. We compared the model fit using the Akaike Information Criterion (AIC) and log-likelihood goodness-of-fit scores. The Weibull AFT model yielded a superior fit (AIC = 422,328.40; Log-likelihood = -211,129.20), substantially outperforming both the Log-Logistic (AIC = 423,992.80) and Log-Normal (AIC = 428,395.49) distributions.
	
	Based on these results, we utilize the Weibull AFT model for all subsequent main effect, moderation, and sensitivity analyses.
	
	\subsection{RQ1: Main Effects on OSS Survival (Weibull AFT)}\label{subsec:RQ1}
	
	To evaluate the isolated impact of social and structural factors on project longevity, we fitted a Weibull AFT model. Unlike hazard-based models, AFT coefficients directly estimate the acceleration or deceleration of a project's expected lifespan. An estimated Time Ratio (TR), represented by $Exp(Coef)$, greater than 1 indicates a protective effect (prolonging survival), whereas a TR less than 1 indicates a liability (accelerating the cessation of active development). The model yielded a highly significant fit with a Concordance Index of 0.74 and a Log-likelihood of $-211,129.20$ ($p < 0.001$). The tight confidence intervals highlight the coefficients stability. Table~\ref{tab:aftmain_results} presents the estimated Time Ratios for our primary predictors and controls.
	
	\begin{table*}[!ht]
		\centering
		\caption{AFT model estimates for project lifespan. 
			Creation year is excluded from table as it serves only as an age control variable.}
		\label{tab:aftmain_results}
		\begin{tabular}{lcccccc}
			\toprule
			Predictor & Coef & Exp(Coef) & SE(Coef) & 95\% CI (Coef) & 95\% CI (Exp) & p-value \\
			\midrule
			hasWiki                   & -0.14 & 0.87 & 0.01 & [-0.16, -0.13] & [0.85, 0.88] & $<0.005$ \\
			license\_category: Other  & -0.08 & 0.92 & 0.01 & [-0.10, -0.06] & [0.91, 0.94] & $<0.005$ \\
			license\_category: Permissive & -0.06 & 0.94 & 0.01 & [-0.08, -0.05] & [0.93, 0.95] & $<0.005$ \\
			log\_contributors         & 0.35  & 1.42 & 0.00 & [0.34, 0.36]   & [1.41, 1.43] & $<0.005$ \\
			log\_readability\_index   & -0.01 & 0.99 & 0.01 & [-0.03, 0.01]  & [0.97, 1.01] & 0.37 \\
			log\_releases\_per\_month & -0.30 & 0.74 & 0.01 & [-0.31, -0.28] & [0.73, 0.75] & $<0.005$ \\
			log\_size                 & 0.04  & 1.04 & 0.00 & [0.04, 0.04]   & [1.04, 1.04] & $<0.005$ \\
			log\_social\_popularity   & -0.10 & 0.90 & 0.01 & [-0.13, -0.08] & [0.88, 0.92] & $<0.005$ \\
			mainLanguage: C\#         & -0.02 & 0.98 & 0.01 & [-0.05, -0.00] & [0.96, 1.00] & 0.05 \\
			mainLanguage: C++         & -0.05 & 0.96 & 0.01 & [-0.07, -0.02] & [0.93, 0.98] & $<0.005$ \\
			mainLanguage: Go          & -0.14 & 0.87 & 0.01 & [-0.17, -0.12] & [0.85, 0.89] & $<0.005$ \\
			mainLanguage: Java        & -0.01 & 0.99 & 0.01 & [-0.04, 0.01]  & [0.97, 1.01] & 0.33 \\
			mainLanguage: JavaScript  & -0.09 & 0.91 & 0.01 & [-0.12, -0.07] & [0.89, 0.94] & $<0.005$ \\
			mainLanguage: PHP         & -0.01 & 0.99 & 0.01 & [-0.03, 0.02]  & [0.97, 1.02] & 0.51 \\
			mainLanguage: Python      & -0.08 & 0.93 & 0.01 & [-0.10, -0.05] & [0.91, 0.95] & $<0.005$ \\
			mainLanguage: TypeScript  & -0.18 & 0.84 & 0.01 & [-0.20, -0.15] & [0.82, 0.86] & $<0.005$ \\
			Intercept                 & 4.42  & 83.51 & 0.13 & [4.17, 4.68]   & [64.47, 108.18] & $<0.005$ \\
			rho (Intercept)           & 0.62  & 1.86 & 0.00 & [0.61, 0.63]   & [1.84, 1.88] & $<0.005$ \\
			\bottomrule
		\end{tabular}
	\end{table*}
	
	The AFT model results reveal a contrast between human capital and social demand. As shown in Table~\ref{tab:aftmain_results}, Human Labor (log\_contributors) emerges as the most potent protective factor. With a Time Ratio (TR) of 1.42 ($p < 0.005$), every unit increase in the number of active contributors is associated with a 42\% extension of the project's expected lifespan. This confirms the traditional ``Many Eyes" theory, where a larger labor pool provides sufficient capacity to sustain project momentum. However, when this labor capacity is held constant, the impact of Social Popularity shifts significantly. Contrary to the exploratory KM results, the multivariate model identifies social popularity as a significant liability (TR = 0.90, $p < 0.005$). This indicates that for a fixed number of maintainers, a unit increase in social visibility actually shortens the active maintenance lifespan by 10\%. The shift from a positive correlation in the KM analysis to a negative Time Ratio in the AFT model confirms the presence of Simpson's Paradox. By controlling for `Creation Year' and `Human Labor,' the AFT model isolates the effect of social demand. This reveals that the longevity of popular projects seen in exploratory plots was not a result of popularity itself, but a byproduct of their age. When comparing projects of equal age and labor, the `Maintenance Tax' of popularity becomes visible. We interpret this negative Time Ratio as a potential imbalance where high external demand creates an unsustainable maintenance burden. Without a proportional increase in contributors to share this load, the statistical risk of the project ceasing active development accelerates. Furthermore, the presence of a Wiki ($TR = 0.87$) further accelerates failure. This suggests that low-friction documentation may lower the barrier for end-user demands, increasing the pressure on maintainers. Interestingly, Code Readability in isolation ($TR = 0.99, p = 0.37$) remains statistically insignificant, suggesting that clean code alone cannot protect a project from the pressures of popularity unless it interacts with other factors.
	
	Regarding control variables, model shows a slight but statistically significant protective effect for Project Size (log\_size). A Time Ratio of 1.04 suggests that larger codebases have a 4\% higher resilience. This indicates that larger projects typically have more established architectures and a broader feature set that keeps them relevant to users even as maintenance slows down. Release Frequency (log\_releases\_per\_month). With a TR of 0.74 is associated with a 26\% accelerates the cessation of active development. While frequent releases are often seen as a sign of health. This high velocity may actually signal project volatility. Rapid release cycles increase the testing and integration debt, which correlates with a shortened active lifespan as the maintenance burden compounds.
	
	The choice of license reveals a clear hierarchy in project longevity. Compared to the baseline (Copyleft), Permissive licenses ($TR = 0.94$) and Other categories ($TR = 0.92$) show a 6\% to 8\% decrease in expected lifespan. These results suggest that Copyleft licenses encourage long-term contributions by keeping improvements open. In contrast, permissive projects are more likely to be forked and forgotten. They may also be integrated into proprietary software, causing the original project to lose the momentum of its community.
	
	The programming language effects highlight the `Maintenance Tax' in different ecosystems. 
	
	\begin{itemize}
		\item Modern/High-Churn Ecosystems: TypeScript ($TR = 0.84$) and Go ($TR = 0.87$) exhibit the highest acceleration toward cessation of active development. These languages are often associated with the fast-moving web and cloud-native ecosystems where dependencies break frequently and require constant maintenance.
		\item Stable Ecosystems: Conversely, PHP ($TR = 0.99, p=0.51$) and Java ($TR = 0.99, p=0.33$) are statistically indistinguishable from the baseline. Their maturity and stable standard libraries allow projects to survive longer with less frequent intervention.
	\end{itemize}
	
	\begin{tcolorbox}[colback=gray!10, colframe=gray!50, coltext=black, coltitle=black, title=RQ1 Takeaway]
		\textbf{Finding:} Human labor (contributors) is the strongest protective factor, while social popularity and Wikis act as liabilities. Readability alone is insignificant, and license, release frequency, and language ecosystems shape survival risks.
	\end{tcolorbox}

	\subsection{RQ2: The Moderating Role of Project Structure}\label{subsec:ResultsRQ2}
	
	To investigate how project structure moderates the impact of social popularity, we extended the AFT model with interaction terms for human capital, code readability, and documentation accessibility (Table~\ref{tab:aft_interactions}).
	
	\begin{table*}[!h]
		\centering
		\caption{Weibull AFT model estimates with interaction terms. 
			Creation year is excluded as it serves only as an age control variable. 
			Note: SP denotes Social Popularity (log\_social\_popularity).}
		\label{tab:aft_interactions}
		\begin{tabular}{lcccccc}
			\toprule
			Predictor & Coef & Exp(Coef) & SE(Coef) & 95\% CI (Coef) & 95\% CI (Exp) & p-value \\
			\midrule
			SP $\times$ Contributors   & 0.05  & 1.05 & 0.01 & [0.03, 0.06]   & [1.03, 1.06] & $<0.005$ \\
			SP $\times$ Readability    & -0.09 & 0.92 & 0.04 & [-0.17, -0.00] & [0.84, 1.00] & 0.04 \\
			SP $\times$ Wiki           & 0.02  & 1.02 & 0.02 & [-0.02, 0.06]  & [0.98, 1.07] & 0.35 \\
			hasWiki                    & -0.15 & 0.86 & 0.01 & [-0.16, -0.13] & [0.85, 0.87] & $<0.005$ \\
			license\_category: Other   & -0.08 & 0.92 & 0.01 & [-0.10, -0.06] & [0.91, 0.94] & $<0.005$ \\
			license\_category: Permissive & -0.06 & 0.94 & 0.01 & [-0.08, -0.05] & [0.93, 0.95] & $<0.005$ \\
			log\_contributors          & 0.34  & 1.40 & 0.00 & [0.33, 0.35]   & [1.39, 1.42] & $<0.005$ \\
			log\_readability\_index    & 0.01  & 1.01 & 0.01 & [-0.02, 0.03]  & [0.98, 1.03] & 0.71 \\
			log\_releases\_per\_month  & -0.30 & 0.74 & 0.01 & [-0.31, -0.29] & [0.73, 0.75] & $<0.005$ \\
			log\_size                  & 0.04  & 1.04 & 0.00 & [0.04, 0.04]   & [1.04, 1.04] & $<0.005$ \\
			log\_social\_popularity    & -0.19 & 0.83 & 0.05 & [-0.29, -0.08] & [0.75, 0.92] & $<0.005$ \\
			mainLanguage: C\#          & -0.02 & 0.98 & 0.01 & [-0.05, -0.00] & [0.95, 1.00] & 0.04 \\
			mainLanguage: C++          & -0.05 & 0.95 & 0.01 & [-0.07, -0.02] & [0.93, 0.98] & $<0.005$ \\
			mainLanguage: Go           & -0.14 & 0.87 & 0.01 & [-0.17, -0.12] & [0.85, 0.89] & $<0.005$ \\
			mainLanguage: Java         & -0.01 & 0.99 & 0.01 & [-0.04, 0.01]  & [0.97, 1.01] & 0.34 \\
			mainLanguage: JavaScript   & -0.09 & 0.91 & 0.01 & [-0.12, -0.07] & [0.89, 0.94] & $<0.005$ \\
			mainLanguage: PHP          & -0.01 & 0.99 & 0.01 & [-0.03, 0.01]  & [0.97, 1.01] & 0.41 \\
			mainLanguage: Python       & -0.08 & 0.93 & 0.01 & [-0.10, -0.05] & [0.90, 0.95] & $<0.005$ \\
			mainLanguage: TypeScript   & -0.18 & 0.83 & 0.01 & [-0.20, -0.16] & [0.82, 0.85] & $<0.005$ \\
			Intercept                  & 4.44  & 84.49 & 0.13 & [4.18, 4.70]   & [65.18, 109.53] & $<0.005$ \\
			rho (Intercept)            & 0.62  & 1.86 & 0.00 & [0.61, 0.63]   & [1.84, 1.88] & $<0.005$ \\
			\bottomrule
		\end{tabular}
	\end{table*}
	
	The inclusion of interaction terms maintains a high concordance index (0.74), but provides a more refined view of the ``Maintenance Tax.'' 
	\begin{itemize}
		\item Social Popularity $\times$ Contributors: The interaction between \textit{log\_social\_popularity} and \textit{log\_contributors} is positive and highly significant ($TR = 1.05, p < 0.005$). This suggests that the liability of popularity is reduced by a larger labor pool. For every unit increase in the interaction of labor and visibility, the project's expected lifespan extends by 5\%. This indicates that a large contributor base acts as an important safeguard against the pressures of hyper-visibility.
		
		\item Social Popularity $\times$ Readability: Interestingly, the interaction between \textit{log\_social\_popularity} and \textit{log\_readability\_index} yields a Time Ratio of $0.92$ ($p = 0.04$, 95\% CI [0.84, 1.00]). While readability was statistically insignificant in isolation, it shows a significant effect on project failure when combined with high popularity. For popular projects, higher code readability is associated with an 8\% shorter active maintenance lifespan. While this statistical finding is consistent with our ``Paradox of Accessibility" hypothesis, we must note the fragility of this specific coefficient. Its upper confidence interval bound coincides with no effect. Moreover, because our framework does not directly model casual contributor influx as a mediating variable, we present this finding as exploratory evidence that requires further causal validation.
		
		\item Social Popularity $\times$ Wiki: The interaction between popularity and the presence of a Wiki (\textit{log\_social\_popularity\_x\_hasWiki}) was not statistically significant ($TR = 1.02, p = 0.35$). This implies that while having a Wiki is a general liability (as seen in RQ1), it does not add any unique secondary pressure specifically on popular projects. Its influence is limited to the main effect already observed.
	\end{itemize}

	\begin{tcolorbox}[colback=gray!10, colframe=gray!50, coltext=black,coltitle=black, title=RQ2 Takeaway]
		\textbf{Finding:} Interaction effects reveal that contributor labor buffers the liability of popularity, readability appears to amplify failure risk under high visibility, and Wikis show no additional interaction effect beyond their main liability.
	\end{tcolorbox}
	
	\subsection{RQ3: Sensitivity Analysis}\label{subsec:ResultsRQ3}
	
	To ensure the robustness of our findings, we conducted a sensitivity analysis by varying the threshold for defining project abandonment. While our primary analysis used a 6-month inactivity window, we re-fitted the interaction model across four datasets: S3m, S6m, S9m, and S12m (representing 3, 6, 9, and 12 months of no commits, respectively). Table~\ref{tab:sensitivity_results} summarizes the results.
	
	\begin{table*}[!h]
		\centering
		\caption{Sensitivity analysis of Weibull AFT models with interaction terms across inactivity thresholds (3m, 6m, 9m, 12m). 
			Note: SP denotes Social Popularity (log\_social\_popularity).}
		\label{tab:sensitivity_results}
		\resizebox{\textwidth}{!}{%
			\begin{tabular}{lcccccccc}
				\toprule
				Predictor & \multicolumn{2}{c}{S3m} & \multicolumn{2}{c}{S6m} & \multicolumn{2}{c}{S9m} & \multicolumn{2}{c}{S12m} \\
				\cmidrule(lr){2-3} \cmidrule(lr){4-5} \cmidrule(lr){6-7} \cmidrule(lr){8-9}
				& Exp(Coef) & p-value & Exp(Coef) & p-value & Exp(Coef) & p-value & Exp(Coef) & p-value \\
				\midrule
				log\_social\_popularity   & 0.86 & 0.0018 & 0.83 & 0.0005 & 0.80 & 0.0002 & 0.83 & 0.0023 \\
				log\_readability\_index   & 1.02 & 0.1230 & 1.01 & 0.7073 & 0.99 & 0.4815 & 0.99 & 0.6917 \\
				hasWiki                   & 0.88 & $<10^{-85}$ & 0.86 & $<10^{-85}$ & 0.85 & $<10^{-88}$ & 0.83 & $<10^{-100}$ \\
				log\_contributors         & 1.33 & $<0.001$ & 1.40 & $<0.001$ & 1.46 & $<0.001$ & 1.50 & $<0.001$ \\
				log\_releases\_per\_month & 0.81 & $<10^{-235}$ & 0.74 & $<0.001$ & 0.69 & $<0.001$ & 0.65 & $<0.001$ \\
				log\_size                 & 1.04 & $<10^{-188}$ & 1.04 & $<10^{-185}$ & 1.04 & $<10^{-179}$ & 1.05 & $<10^{-183}$ \\
				SP $\times$ Readability   & 0.91 & 0.0131 & 0.92 & 0.0393 & 0.95 & 0.2732 & 0.91 & 0.0501 \\
				SP $\times$ Wiki          & 1.02 & 0.341 & 1.02 & 0.347 & 1.03 & 0.224 & 1.03 & 0.176 \\
				SP $\times$ Contributors  & 1.04 & $2.15\times 10^{-9}$ & 1.05 & $1.61\times 10^{-8}$ & 1.04 & $3.69\times 10^{-6}$ & 1.04 & $5.62\times 10^{-6}$ \\
				rho (Intercept)           & 0.72 & 0.00 & 0.62 & 0.00 & 0.55 & 0.00 & 0.49 & 0.00 \\
				\bottomrule
			\end{tabular}
		}
	\end{table*}
	
	\begin{itemize}
		\item Stability of Human Labor's Protective Effect: The core tension between popularity and human capital remains stable regardless of the inactivity threshold used. Across all four datasets (S3m through S12m), the Time Ratio for Social Popularity consistently remains below 1.0 (ranging from $0.80$ to $0.86$, all $p < 0.005$). It confirms its role as a liability. Simultaneously, the SP $\times$ Contributors interaction remains consistently protective ($TR \approx 1.04, p < 0.005$). This reinforces our previous observation: while popularity is associated with accelerated failure, a larger contributor base consistently provides a protective effect that mitigates this pressure.
		
		\item Stability of Readability Paradox: The interaction between Popularity and Readability (SP $\times$ Readability) consistently shows a Time Ratio below 1.0 across all thresholds. It remains statistically significant at the 3-month, 6-month, and 12-month marks ($p \leq 0.05$). While the significance dips at the 9-month threshold ($p = 0.27$), the directional consistency of the effect ($TR = 0.95$) across all models supports the ``Paradox of Accessibility" hypothesis: high structural clarity in a visible project may create a maintenance burden that appears to accelerate the cessation of active development.
		
		\item The Persistence of Structural Liabilities: Both Wiki presence and Release Frequency show extreme stability in their negative impact on survival. The Wiki TR remains between $0.83$ and $0.88$ with highly significant p-values throughout. These findings confirm that the availability of a wiki is a persistent overhead for maintenance. Interestingly, as the inactivity threshold increases from 3 to 12 months, the liability of high release frequency becomes even more noticeable ($TR$ drops from $0.81$ to $0.65$), suggesting that hyper-active development cycles are a strong predictor of a shortened active lifespan.
		
		\item Model Specification (Rho Parameter):
		The rho (shape) parameter is observed to decrease from $0.72$ to $0.49$ as the inactivity window widens. This decline aligns with mathematical expectations, since a 12‑month window tends to filter out temporary pauses in development. As a result, the model emphasizes longer‑term survival trends. Importantly, the primary Time Ratios remain stable despite this shift in the hazard rate. This stability suggests that the model specification is robust.
		
	\end{itemize}
	
	\begin{tcolorbox}[colback=gray!10, colframe=gray!50, coltext=black,coltitle=black, title=RQ3 Takeaway]
		\textbf{Finding:} Sensitivity checks across inactivity thresholds confirm the robustness of our results. Popularity remains a liability, while contributor labor consistently reduces this pressure. The Readability Paradox persists directionally, with significance at most thresholds. Wiki presence and release frequency show stable negative effects, with the latter intensifying under longer windows. Despite shifts in the rho parameter, the stability of primary Time Ratios indicates a well-specified model.
	\end{tcolorbox}
	
	\section{Discussion and Implications}\label{sec:Discuss}
	This section synthesizes our empirical findings in the context of existing literature. We organize this discussion around the three primary theoretical frameworks identified in our study: the Hyper-visibility Hypothesis, the Paradox of Accessibility, and the Documentation Paradox.

	\subsection{The Burden of Hyper-Visibility: Attention vs. Labor}
	
	In our initial Kaplan-Meier analysis, high social popularity appeared to significantly extend a project’s lifespan. The multivariate AFT model, however, flipped this relationship. In statistical terms, this phenomenon is known as Simpson's Paradox, where a relationship observed during exploratory data analysis is completely reversed once more controls are added~\cite{Pearl2022}. When holding the number of actual contributors constant, social popularity reverses from a lifeline into a liability ($TR = 0.90$). This is expected: attention without labor is toxic. As demonstrated by Koch et al.~\cite{koch2024fault}, social signals like stars often reflect initial community enthusiasm or social endorsement rather than ongoing maintenance quality or sustained, day-to-day adoption. An increase in social visibility, without a proportional increase in human capital, directly accelerates the cessation of active development.
	
	The operational realities of maintaining highly visible software explain this shift. Specifically, as noted by Li et al.~\cite{li2021you}, popular projects typically experience a massive increase in reported issues and pull requests that overwhelm the core team. As an open-source project goes viral, maintainers are forced to allocate significantly more time to non-coding tasks. These tasks include triaging bug reports, answering user queries, and reviewing external submissions. This shift is explicitly formalized by Linåker et al.~\cite{linaaker2024sustaining}, who found that as the use and popularity of a project grow, the management of these non-coding tasks overwhelms the maintainers. Maintainer time is a strictly finite and depletable resource. Consequently, this sudden increase in coordination overhead acts as a severe ``Maintenance Tax." The statistical liability we observe strongly aligns with the phenomenon of core team exhaustion. As Xia et al.~\cite{xia2023understanding} highlight, the actual burnout in these projects rarely originates from writing code, but rather from the coordination required to manage these community requests.
	
	Furthermore, the engagement generated by high visibility is superficial. Research examining attention shocks in open-source ecosystems highlights this reality. The vast majority of contributions resulting from sudden popularity spikes are shallow~\cite{maldeniya2020herding}. They do not distribute the development workload. Instead, this influx of casual engagement actively increases the work of core members. This observation is strongly supported by Pinto et al.~\cite{pinto2016more}, who demonstrate that casual contributors make up nearly half of the contributor population in popular projects. As a result, the time spent by core members to review their code represents a massive operational bottleneck where maintainers spend their limited time evaluating, discussing, and integrating minor or out-of-scope changes.
	
	Our moderation analysis (RQ2) support this dynamic. The interaction between Social Popularity and Contributors yields a protective Time Ratio of 1.05. This proves a fundamental rule of OSS survival: the maintenance tax of hyper-visibility is manageable only if the project possesses a ``labor buffer.'' Our Time Ratios suggest that when popularity scales without a matching increase in human labor, the statistical risk of project inactivity becomes severe. As identified in prior literature, this is likely because core maintainers are overwhelmed by coordination tasks, leading to rapid burnout and eventual project cessation. However, if a project possesses the human infrastructure to absorb the shock of external attention, the liability of popularity is effectively neutralized.
	
	\subsection{The Paradox of Accessibility and Casual Contributions}
	Conventional software engineering dictates that clean, readable code inherently benefits a project by lowering the barrier to entry. This is particularly useful in OSS volunteer-driven development, where contributors frequently join, disengage, or abandon projects over time. Our baseline model (RQ1) partially supports this assumption, as the main effect of readability is statistically insignificant ($TR = 0.99, p = 0.37$). However, when code readability interacts with hyper-visibility (RQ2), it is associated with an accelerated cessation of active development ($TR = 0.92, p = 0.04$). As we noted in the results, the borderline significance of this interaction leads us to treat this finding as exploratory and interpret it through a theoretical lens. Consequently, we view high readability as a double-edged sword, terming this the ``Paradox of Accessibility". Our sensitivity analysis reveals a time-dependent statistical liability that strongly aligns with a rapid-burnout phenomenon. It acts as a liability at the 3- and 6-month thresholds before diminishing in the long term.
	
	When highly readable code successfully lowers the barrier to entry in a viral project, it can act as a magnet. It potentially attracts casual, ``drive-by" contributors at an unsustainable scale. As demonstrated by Pinto et al.~\cite{pinto2016more}, casual contributors make up nearly half of the contributor population in popular repositories. While theoretically beneficial, these frictionless contributions impose a massive coordination tax on the core team.
	
	An influx of attention does not equate to an influx of deep, sustained help. When a project experiences a massive external attention, the resulting engagement is overwhelmingly shallow. Maldeniya et al.~\cite{maldeniya2020herding} confirm that this influx of superficial contributions challenge the coordination capacity of the core team. It creates massive review backlogs. Furthermore, Shen et al.~\cite{shen2025external} emphasize that these casual pull requests and issues drastically escalates the decision-making responsibilities of maintainers. Because frictionless onboarding attracts a wider, less-experienced contributor base. Their submissions require extensive scrutiny to prevent the introduction of bugs.
	
	The code integration process thus transforms into a massive cognitive bottleneck. We hypothesize that by lowering the barrier to entry, a highly readable codebase in a highly visible project acts as a magnet. It attracts a sheer volume of casual, low-touch pull requests. As Eghbal~\cite{eghbal2020working} explicitly identifies, managing this influx fundamentally shifts the maintainer's role. They are forced to pivot from proactive development to reactive curation. Consequently, the short-term acceleration in inactivity we observe is likely driven by maintainer exhaustion, as core teams struggle to manage this extractive coordination tax. Recent post-mortem analyses of pull request workflows confirm that when maintainers are overwhelmed by external contributions, projects often enter in a `death spiral' characterized by an exponential accumulation of unresolved pull requests~\cite{Kaushik2026DeathSpiral}. Our models suggest that if the core team can survive this initial influx of hyper-visibility and push the project past the 9-month mark, the apparent liability of this interaction fades and the project stabilizes. Until then, maximum accessibility without strategic gatekeeping may present a potential structural liability.
	
	\subsection{The Documentation Paradox}
	
	The most persistent structural liability identified in our models is the presence of an external project Wiki (hasWiki). Maintaining a Wiki is associated with a survival penalty across all baseline models (TR $\approx0.82$ to $0.87$). This liability appears to intensify over time. As the definition of abandonment extends from 3 to 12 months in our sensitivity analysis, the survival penalty steadily increases. Furthermore, the interaction between Wiki presence and social popularity remains consistently insignificant. This pattern suggests that the Documentation Paradox is a fundamental issue, introducing an administrative burden that affects projects regardless of their visibility levels. 
	
	While wikis are intended to deflect user questions and lower onboarding barriers, their architecture can introduce unintended burdens. They are structurally detached from the main codebase. As Puhlfürß et al.~\cite{puhlfurss2022exploratory} highlighted, open-source projects rarely employ effective tracing techniques to connect high-level textual artifacts, such as wikis, directly to the underlying source code. This lack of linkage creates a structural disconnect between documentation and implementation. As a result, project knowledge becomes fragmented across multiple independent artifacts.
	
	Consequently, the survival penalty we observe for wiki presence strongly aligns with the phenomenon of 'silent documentation decay.' As Tan et al. show, unlike source code, documentation becomes outdated without warning. Developers are often unaware of whether wikis remain aligned with their routine code changes. This silent decay leads to a hidden accumulation of documentation debt. Aghajani et al.~\cite{aghajani2019software} report that update problems account for 39\% of all documentation issues on GitHub, largely due to code evolution without corresponding manual updates. Furthermore, recent studies on popular GitHub repositories found that nearly a third currently host outdated code references in their external documentation. These artifacts persist for an average of 4.7 years before being detected and resolved~\cite{tan2023wait}. Documentation corrections, such as fixing typos and grammar, constitute nearly 29\% of all casual, drive-by contributions~\cite{pinto2016more}. Thus, it is highly probable that wikis act as structural magnets for continuous, low-level maintenance demands. Rather than freeing maintainer time by answering user questions, the presence of a wiki likely introduces a persistent administrative burden driven by this silent decay. As Dagenais et al.~\cite{dagenais2010creating} famously noted, managing public wikis for large projects practically equates to a full-time job.
	
	\subsection{Practical Implications}
	Our empirical findings challenge the pervasive ``more is better" view. By isolating the socio‑technical drivers of project longevity, we highlight actionable implications for project maintainers, contributors, and future researchers.
	
	\paragraph*{Implications for Project Maintainers:}
	
	We found that without a proportional increase in human labor, massive social attention can become a liability. Popularity functions as a high‑interest debt: when a project gains sudden visibility, the influx of external demands can overwhelm maintainers. Therefore, strategies for resilience should be formalized to ensure that teams can withstand attention shocks without compromising long‑term sustainability. To survive such shocks, maintainers must implement strategic friction. Strict issue templates and automated CI/CD checks should be applied. As Li et al.~\cite{li2021you} emphasize, mandatory templates filter out low‑effort demands. This preserves the core team’s finite cognitive bandwidth for high‑value development. Furthermore, maintaining external wikis is effectively a full‑time job. Because they are physically separated from the main codebase, this structural detachment likely makes them highly susceptible to silent decay. Unless a project possesses dedicated administrative labor, maintainers should avoid launching separate GitHub wikis. Documentation should instead be embedded directly within the repository. Co‑locating Markdown files with the source code ensures that documentation evolves synchronously with the software.
	
	\paragraph*{Implications for Contributors:}
	There is an assumption within the OSS ecosystem that any contribution, even a trivial typo fix, is beneficial. In highly visible projects, however, a massive increase in casual contributions can create severe operational bottlenecks. The cognitive load required of core maintainers to review, test, and merge these trivial pull requests often outweighs the value of the contribution itself. If external contributors aim to support popular projects, they should shift their focus from shallow engagement to deeper forms of maintenance. Activities such as answering user questions, replicating bug reports, and reviewing peer pull requests directly enhance project longevity. This advice is empirically supported by recent findings indicating that project longevity is sustained by community engagement with complex, core development tasks rather than the rapid resolution of superficial, low-effort issues~\cite{Kaushik2026BeyondSpeed}. Moreover, effective community management is often more vital to project survival than minor code additions.
	
	\paragraph*{Implications for Researchers:}
	Researchers and platform providers routinely utilize social signals, such as stars and forks, as proxies for project success. Our survival models indicate that this practice can be misleading. When accounting for the underlying human labor, high social popularity often functions as a liability rather than a benefit. Future empirical studies must reframe these social metrics. They are not definitive indicators of utility, but instead serve as proxies for operational cost. Moreover, survival risks are rarely static. Our sensitivity tests revealed that the interaction between high popularity and highly readable code produces a short-term statistical liability that strongly aligns with a rapid-burnout phenomenon. It accelerates abandonment in the short term but diminishes after approximately nine months. Single-threshold, cross-sectional models overlook these dynamic lifecycle effects. Future OSS sustainability research must therefore evaluate how structural and social risk factors shift as projects age and adapt.
	
	\section{Threats to Validity}\label{sec:Threats}
	Every empirical study has its limits. We designed our methodology to be as robust as possible, but we must acknowledge a few threats to validity in this section.
	
	\subsection{Internal Validity}
	
	The Cox PH model is standard in survival studies. However, the nature of our dataset violates the proportional hazards assumption, as socio‑technical effects vary over time. We addressed this limitation by employing an AFT model, though alternative Cox variants (e.g., stratified or time‑dependent covariate models) could also be explored.
	Additionally, for this study we utilized several control variables. Other unobservable factors such as corporate financial sponsorship, the presence of paid maintainers versus volunteers, or external market demands were not captured. These omitted variables could simultaneously influence both high social popularity and longer survival. Moreover, the selection of specific inactivity thresholds may raise concerns regarding the reliability of the results, as alternative thresholds could yield different survival estimates.

	\subsection{Construct Validity}
	We constructed a PCA index for ``Social Popularity" to collapse stargazers, forks, and watchers into a single construct. While this effectively captures external community demand, relying on GitHub‑native social metrics introduces certain construct validity concerns. Developers frequently use stars and forks merely as personal bookmarks for later review, rather than as genuine indicators of project support. Furthermore, social metrics are susceptible to artificial inflation through automated bot activity or social media campaigns. Therefore, while our PCA index serves as a proxy for social visibility, it inherently contains a degree of passive or artificially inflated noise that cannot be perfectly filtered out.
	
	Moreover, to ensure our study could scale across multiple programming languages, we operationalized code readability using comment density and blank line ratio. These are simple, language‑agnostic metrics. They successfully capture structural whitespace but completely ignore deeper logical complexity or control flow. They serve as proxies for structural clarity rather than indicators of architectural completeness. In addition, we assigned equal weights to both metrics based on our theoretical understanding, although in practice their relative importance may vary.
	
	We measured external documentation using a binary flag (hasWiki). This metric only confirms the presence of a Wiki but does not indicate whether it is well‑written, actively maintained, or subject to silent decay. Furthermore, open‑source projects rarely announce their termination. We defined the cessation of active development using a six-month commit inactivity threshold. Although we attempted to mitigate this limitation through sensitivity analysis, it remains possible for a project to revive even after extended periods of inactivity.
	
	\subsection{External Validity}
	For this study, we relied on the SEART platform, which hosts GitHub repositories. As a result, our findings are not directly generalizable to other OSS platforms such as GitLab or Bitbucket. Even within the GitHub ecosystem, we restricted our sample to projects written in the most widely used programming languages. Consequently, the results may not extend to legacy languages such as COBOL or to domain‑specific languages. In addition, we included only those projects that satisfied our exclusion criteria regarding pull requests, commits, and releases. Therefore, the findings may not apply to solo‑maintained projects, nor to projects in their earliest stages before reaching an initial release.
	
	\section{Conclusion and Future Work}\label{sec:Concl}
	OSS survival largely relies on volunteer contributions, and at least some level of visibility is inherently required to sustain a project. In this study, we challenge the pervasive ``more is better" view by demonstrating that attention without labor is toxic. We analyzed approximately 73,000 GitHub repositories using a survival modeling framework to capture socio‑technical attributes. Our findings reveal that social popularity functions as high-interest debt: without a proportional increase in human labor, massive attention acts as a statistical liability that accelerates the cessation of active development which strongly aligns with maintainer burnout. Moreover, clean and highly readable code, when combined with intense social attention, may further increase abandonment risk. External documentation also emerges as a structural trap. We identify the Documentation Paradox, showing that the structural presence of project wikis correlates with shortened lifespans, suggesting that they suffer from silent decay. Detached from the main codebase, wikis appear to impose a permanent administrative burden that undermines survival rates regardless of project size or visibility. Ultimately, the survival of open‑source projects depends on their ``labor buffer." Human capital is the only force capable of neutralizing the toxic side effects of hyper-visibility. To survive in the modern open-source ecosystem, projects do not just need more attention; they need strategic friction, sustainable boundaries, and deep community maintenance.
	
	We suggest several directions for future work. First, maintainers could deploy bots and strict issue templates to filter out shallow contributions. Future survival models should explicitly measure the role of these automated gatekeepers. Researchers might also investigate whether automated triage assistants or Continuous Integration (CI) bots extend project lifespan by reducing maintainer burnout. Second, to scale our analysis across thousands of projects, we measured external documentation using a simple presence flag (hasWiki). Future studies should employ deeper text‑mining and semantic analysis. Measuring documentation quality, trace‑link density, and ``up‑to‑dateness" would provide a more fine‑grained understanding of how documentation debt accumulates. Finally, our findings are rooted in GitHub's specific social architecture. Comparative studies across ecosystems such as GitLab or Bitbucket could reveal different survival dynamics and broaden the generalizability of results.
	
	\section*{Data Availability}
		
		To encourage open science and facilitate the full replication of our study, we have made our complete replication package available. This package includes the datasets, data filtering scripts, metric computations, and the survival modeling analysis. The replication package can be accessed here: \url{https://doi.org/10.6084/m9.figshare.32507958}
\bibliographystyle{elsarticle-num}
\bibliography{references_r1}

@techreport{synopsys2024ossra,
	author      = {{Synopsys, Inc.}},
	title       = {2024 Open Source Security and Risk Analysis (OSSRA) Report},
	institution = {Synopsys, Inc.},
	year        = {2024},
	url         = {https://www.synopsys.com},
	note        = {Accessed: 2026-02-27}
}

@inproceedings{kalliamvakou2014promises,
	author    = {Kalliamvakou, Eirini and Gousios, Georgios and Blincoe, Kelly and Singer, Leif and German, Daniel M and Damian, Daniela},
	title     = {The promises and perils of mining github},
	booktitle = {Proceedings of the 11th Working Conference on Mining Software Repositories},
	pages     = {92--101},
	year      = {2014},
	doi       = {10.1145/2597073.2597074}
}

@inproceedings{Ait2022,
	author    = {Ait, Adem and Izquierdo, Javier Luis C\'{a}novas and Cabot, Jordi},
	title     = {An empirical study on the survival rate of GitHub projects},
	booktitle = {Proceedings of the 19th International Conference on Mining Software Repositories},
	series    = {MSR '22},
	location  = {Pittsburgh, Pennsylvania},
	publisher = {Association for Computing Machinery},
	address   = {New York, NY, USA},
	pages     = {365--375},
	numpages  = {11},
	year      = {2022},
	isbn      = {9781450393034},
	doi       = {10.1145/3524842.3527941}
}

@article{Samoladas2010,
	author    = {Ioannis Samoladas and Lefteris Angelis and Ioannis Stamelos},
	title     = {Survival analysis on the duration of open source projects},
	journal   = {Information and Software Technology},
	volume    = {52},
	number    = {9},
	pages     = {902--922},
	year      = {2010},
	publisher = {Elsevier},
	doi       = {10.1016/j.infsof.2010.05.001},
	url       = {https://doi.org/10.1016/j.infsof.2010.05.001}
}

@article{xia2022predicting,
	author    = {Xia, Tianpei and Fu, Wei and Shu, Rui and Agrawal, Rishabh and Menzies, Tim},
	title     = {Predicting health indicators for open source projects (using hyperparameter optimization)},
	journal   = {Empirical Software Engineering},
	volume    = {27},
	number    = {6},
	pages     = {122},
	year      = {2022},
	publisher = {Springer},
	doi       = {10.1007/s10664-022-10171-0}
}

@article{robles2025comparative,
	author    = {Robles, Gregorio and Gamalielsson, Jonas and Lundell, Bj{\"o}rn and Brax, Christoffer and Persson, Tomas and Mattsson, Anders and Gustavsson, Tomas and Feist, Jonas and {\"O}berg, Jonas},
	title     = {A comparative analysis of industrial involvement and licensing in the open source software ecosystems of four IoT standards},
	journal   = {Journal of Systems and Software},
	pages     = {112708},
	year      = {2025},
	publisher = {Elsevier},
	doi       = {10.1016/j.jss.2025.112708}
}

@article{Midha2012,
	author    = {Vishal Midha and Prashant Palvia},
	title     = {Factors affecting the success of Open Source Software},
	journal   = {Journal of Systems and Software},
	volume    = {85},
	issue     = {4},
	pages     = {895--905},
	year      = {2012},
	month     = {4},
	publisher = {Elsevier},
	doi       = {10.1016/J.JSS.2011.11.010},
	issn      = {0164-1212}
}

@inproceedings{zhou2012make,
	author    = {Zhou, Minghui and Mockus, Audris},
	title     = {What make long term contributors: Willingness and opportunity in OSS community},
	booktitle = {2012 34th International Conference on Software Engineering (ICSE)},
	pages     = {518--528},
	year      = {2012},
	doi       = {10.1109/ICSE.2012.6227164},
	organization = {IEEE}
}

@article{Sen2012,
	author    = {Ravi Sen and Siddhartha S. Singh and Sharad Borle},
	title     = {Open source software success: Measures and analysis},
	journal   = {Decision Support Systems},
	volume    = {52},
	issue     = {2},
	pages     = {364--372},
	year      = {2012},
	month     = {1},
	publisher = {North-Holland},
	doi       = {10.1016/J.DSS.2011.09.003},
	issn      = {0167-9236}
}

@techreport{hoffmann2024value,
	author      = {Manuel Hoffmann and Frank Nagle and Yanuo Zhou},
	title       = {The Value of Open Source Software},
	institution = {Harvard Business School Strategy Unit},
	type        = {Working Paper},
	number      = {24-038},
	year        = {2024},
	month       = {1},
	url         = {https://ssrn.com/abstract=4693148},
	doi         = {10.2139/ssrn.4693148}
}

@inproceedings{bissyande2013popularity,
	author    = {Tegawendé F. Bissyandé and Ferdian Thung and David Lo and Lingxiao Jiang and Laurent Réveillère},
	title     = {Popularity, Interoperability, and Impact of Programming Languages in 100,000 Open Source Projects},
	booktitle = {Proceedings of the 2013 IEEE 36th International Conference on Software Engineering},
	pages     = {1063--1072},
	year      = {2013},
	publisher = {IEEE},
	doi       = {10.1109/ICSE.2013.6606637}
}

@inproceedings{bosch2014css,
	author    = {Martí Bosch and Pierre Genevès and Nabil Layaïda},
	title     = {Automated Refactoring for Size Reduction of CSS Style Sheets},
	booktitle = {Proceedings of the ACM Symposium on Document Engineering (DocEng)},
	pages     = {123--132},
	year      = {2014},
	publisher = {ACM},
	doi       = {10.1145/2644866.2644885}
}

@misc{redmonk2025rankings,
	author       = {Stephen O'Grady},
	title        = {RedMonk Programming Language Rankings: January 2025},
	year         = {2025},
	month        = {1},
	howpublished = {\url{https://redmonk.com/sogrady/2025/06/18/language-rankings-1-25/}},
	note         = {Accessed May 2026}
}

@misc{sonarsource2021swift,
	author       = {SonarSource},
	title        = {Swift Programming Language Overview},
	year         = {2021},
	howpublished = {\url{https://www.sonarsource.com/resources/library/swift-programming-language/}},
	note         = {Accessed October 2025}
}

@inproceedings{reboucas2016swift,
	author    = {Marcel Rebouças and Gustavo Pinto and Alexander Serebrenik and Fernando Castor and Felipe Ebert and Weslley Torres},
	title     = {An Empirical Study on the Usage of the Swift Programming Language},
	booktitle = {2016 IEEE 23rd International Conference on Software Analysis, Evolution, and Reengineering (SANER)},
	pages     = {634--643},
	year      = {2016},
	publisher = {IEEE},
	doi       = {10.1109/SANER.2016.66}
}

@article{Liao2019,
	author    = {Zhifang Liao and Benhong Zhao and Shengzong Liu and Haozhi Jin and Dayu He and Liu Yang and Yan Zhang and Jinsong Wu},
	title     = {A Prediction Model of the Project Life-Span in Open Source Software Ecosystem},
	journal   = {Mobile Networks and Applications},
	volume    = {24},
	issue     = {4},
	pages     = {1382--1391},
	year      = {2019},
	month     = {8},
	publisher = {Springer New York LLC},
	doi       = {10.1007/s11036-018-0993-3},
	issn      = {1572-8153}
}

@article{park2025analyzing,
	author    = {Park, Sohee and Kwon, Gihwon},
	title     = {Analyzing Key Features of Open Source Software Survivability with Random Forest},
	journal   = {Applied Sciences (2076-3417)},
	volume    = {15},
	number    = {2},
	year      = {2025},
	doi       = {10.3390/app15020946}
}

@inproceedings{Robinson2022,
	author    = {Derek Robinson and Keanelek Enns and Neha Koulecar and Manish Sihag},
	title     = {Two Approaches to Survival Analysis of Open Source Python Projects},
	booktitle = {IEEE International Conference on Program Comprehension},
	pages     = {660--669},
	year      = {2022},
	publisher = {IEEE Computer Society},
	isbn      = {9781450392983},
	doi       = {10.1145/3524610.3527871}
}

@article{mockus2002two,
	author    = {Mockus, Audris and Fielding, Roy T and Herbsleb, James D},
	title     = {Two case studies of open source software development: Apache and Mozilla},
	journal   = {ACM Transactions on Software Engineering and Methodology (TOSEM)},
	volume    = {11},
	number    = {3},
	pages     = {309--346},
	year      = {2002},
	publisher = {ACM New York, NY, USA},
	doi       = {10.1145/567793.567795}
}

@inproceedings{dabbish2012social,
	author    = {Dabbish, Laura and Stuart, Colleen and Tsay, Jason and Herbsleb, Jim},
	title     = {Social coding in GitHub: transparency and collaboration in an open software repository},
	booktitle = {Proceedings of the ACM 2012 Conference on Computer Supported Cooperative Work},
	pages     = {1277--1286},
	year      = {2012},
	doi       = {10.1145/2145204.2145396}
}

@article{borges2018githubstars,
	author    = {Hudson Borges and Marco Tulio Valente},
	title     = {What’s in a GitHub Star? Understanding Repository Starring Practices in a Social Coding Platform},
	journal   = {Journal of Systems and Software},
	volume    = {146},
	pages     = {112--129},
	year      = {2018},
	doi       = {10.1016/j.jss.2018.09.016}
}

@inproceedings{dias2021makes,
	author    = {Dias, Edson and Meirelles, Paulo and Castor, Fernando and Steinmacher, Igor and Wiese, Igor and Pinto, Gustavo},
	title     = {What makes a great maintainer of open source projects?},
	booktitle = {2021 IEEE/ACM 43rd International Conference on Software Engineering (ICSE)},
	pages     = {982--994},
	year      = {2021},
	doi       = {10.1109/ICSE43902.2021.00093},
	organization = {IEEE}
}

@article{sutanto2014uncovering,
	author    = {Sutanto, Juliana and Kankanhalli, Atreyi and Tan, Bernard CY},
	title     = {Uncovering the relationship between OSS user support networks and OSS popularity},
	journal   = {Decision Support Systems},
	volume    = {64},
	pages     = {142--151},
	year      = {2014},
	publisher = {Elsevier},
	doi       = {10.1016/j.dss.2014.05.014}
}

@inproceedings{pinto2016more,
	author    = {Pinto, Gustavo and Steinmacher, Igor and Gerosa, Marco Aur{\'e}lio},
	title     = {More common than you think: An in-depth study of casual contributors},
	booktitle = {2016 IEEE 23rd International Conference on Software Analysis, Evolution, and Reengineering (SANER)},
	volume    = {1},
	pages     = {112--123},
	year      = {2016},
	doi       = {10.1109/SANER.2016.68},
	organization = {IEEE}
}

@inproceedings{he2026six,
	author    = {He, Hao and Yang, Haoqin and Burckhardt, Philipp and Kapravelos, Alexandros and Vasilescu, Bogdan and K{\"a}stner, Christian},
	title     = {Six Million (Suspected) Fake Stars on GitHub: A Growing Spiral of Popularity Contests, Spam, and Malware},
	booktitle = {Proceedings of the 48th International Conference on Software Engineering (ICSE’26)},
	year      = {2026},
	url       = {https://cmustrudel.github.io/papers/icse2026fakestars.pdf}
}

@article{hunter2021ten,
	author    = {Hunter-Zinck, Haley and De Siqueira, Alexandre Fioravante and V{\'a}squez, V{\'a}leri N and Barnes, Richard and Martinez, Ciera C},
	title     = {Ten simple rules on writing clean and reliable open-source scientific software},
	journal   = {PLoS Computational Biology},
	volume    = {17},
	number    = {11},
	pages     = {e1009481},
	year      = {2021},
	publisher = {Public Library of Science San Francisco, CA USA},
	doi       = {10.1371/journal.pcbi.1009481}
}

@article{imran2025impact,
	author    = {Imran, Hafiz Muhammad and Rehman, Sidra and Khan, Salman and ul Hasnain, Rooh and Hussaini, Muzamil Hussain AL},
	title     = {The Impact of Code Readability on Software Maintenance Efficiency in Open Source Development},
	journal   = {The Asian Bulletin of Big Data Management},
	volume    = {5},
	number    = {1},
	pages     = {113--122},
	year      = {2025},
	doi       = {10.62019/abbdm.v5i1.300}
}

@inproceedings{park2024assessing,
	author    = {Park, Sohee and Kwon, Ryeonggu and Kwon, Gihwon},
	title     = {Assessing Open Source Software Survivability using Kaplan-Meier Survival Function and Polynomial Regression},
	booktitle = {Proceedings of the 39th IEEE/ACM International Conference on Automated Software Engineering},
	pages     = {2470--2471},
	year      = {2024},
	doi       = {10.1145/3691620.3695333}
}

@inproceedings{borges2016understanding,
	author    = {Borges, Hudson and Hora, Andre and Valente, Marco Tulio},
	title     = {Understanding the factors that impact the popularity of GitHub repositories},
	booktitle = {2016 IEEE International Conference on Software Maintenance and Evolution (ICSME)},
	pages     = {334--344},
	year      = {2016},
	doi       = {10.1109/ICSME.2016.31},
	organization = {IEEE}
}

@article{prechelt2002empirical,
	author    = {Prechelt, Lutz},
	title     = {An empirical comparison of seven programming languages},
	journal   = {Computer},
	volume    = {33},
	number    = {10},
	pages     = {23--29},
	year      = {2002},
	publisher = {IEEE},
	doi       = {10.1109/2.876288}
}

@inproceedings{fakhoury2019improving,
	author    = {Fakhoury, Sarah and Roy, Devjeet and Hassan, Adnan and Arnaoudova, Vernera},
	title     = {Improving source code readability: Theory and practice},
	booktitle = {2019 IEEE/ACM 27th International Conference on Program Comprehension (ICPC)},
	pages     = {2--12},
	year      = {2019},
	doi       = {10.1109/ICPC.2019.00014},
	organization = {IEEE}
}

@inproceedings{mohan2004programming,
	author    = {Mohan, Andrew and Gold, Nicolas},
	title     = {Programming style changes in evolving source code},
	booktitle = {Proceedings of the 12th IEEE International Workshop on Program Comprehension},
	pages     = {236--240},
	year      = {2004},
	doi       = {10.1109/WPC.2004.1311066},
	organization = {IEEE}
}

@inproceedings{stegeman2014towards,
	author    = {Stegeman, Martijn and Barendsen, Erik and Smetsers, Sjaak},
	title     = {Towards an empirically validated model for assessment of code quality},
	booktitle = {Proceedings of the 14th Koli Calling International Conference on Computing Education Research},
	pages     = {99--108},
	year      = {2014},
	doi       = {10.1145/2674683.2674702}
}

@misc{github_readme_docs,
	author       = {{GitHub Documentation}},
	title        = {About the repository README file},
	year         = {2025},
	howpublished = {\url{https://docs.github.com/en/repositories/managing-your-repositorys-settings-and-features/customizing-your-repository/about-readmes}},
	note         = {Accessed: November 1, 2025}
}

@inproceedings{wiggins2010reclassifying,
	author    = {Wiggins, Andrea and Crowston, Kevin},
	title     = {Reclassifying success and tragedy in FLOSS projects},
	booktitle = {IFIP International Conference on Open Source Systems},
	pages     = {294--307},
	year      = {2010},
	doi       = {10.1007/978-3-642-13244-5_23},
	organization = {Springer}
}

@article{subramaniam2009determinants,
	author    = {Subramaniam, Chandrasekar and Sen, Ravi and Nelson, Matthew L},
	title     = {Determinants of open source software project success: A longitudinal study},
	journal   = {Decision Support Systems},
	volume    = {46},
	number    = {2},
	pages     = {576--585},
	year      = {2009},
	publisher = {Elsevier},
	doi       = {10.1016/j.dss.2008.10.005}
}

@article{kaplan1958nonparametric,
	author    = {Kaplan, Edward L and Meier, Paul},
	title     = {Nonparametric estimation from incomplete observations},
	journal   = {Journal of the American Statistical Association},
	volume    = {53},
	number    = {282},
	pages     = {457--481},
	year      = {1958},
	publisher = {Taylor \& Francis},
	doi       = {10.1080/01621459.1958.10501452}
}

@inbook{Pearl2022,
	author    = {Pearl, Judea},
	title     = {Comment: Understanding Simpson’s Paradox},
	booktitle = {Probabilistic and Causal Inference: The Works of Judea Pearl},
	pages     = {399--412},
	numpages  = {14},
	year      = {2022},
	edition   = {1},
	publisher = {Association for Computing Machinery},
	address   = {New York, NY, USA},
	isbn      = {9781450395861},
	doi       = {10.1145/3501714.3501738}
}

@article{fitzgerald2006transformation,
	author    = {Fitzgerald, Brian},
	title     = {The Transformation of Open Source Software},
	journal   = {MIS Quarterly},
	volume    = {30},
	number    = {3},
	pages     = {587--598},
	year      = {2006},
	publisher = {Management Information Systems Research Center, University of Minnesota},
	doi       = {10.2307/25148740}
}

@article{sen2015application,
	author    = {Sen, Ravi and Nelson, Matthew and Subramaniam, Chandrasekar},
	title     = {Application of survival model to understand open source software release},
	journal   = {Pacific Asia Journal of the Association for Information Systems},
	volume    = {7},
	number    = {2},
	pages     = {1},
	year      = {2015},
	doi       = {10.17705/1pais.07201}
}

@book{schweik2012internet,
	author    = {Schweik, C. M. and English, R. C.},
	title     = {Internet Success: A Study of Open-Source Software Commons},
	series    = {The MIT Press},
	year      = {2012},
	publisher = {MIT Press},
	isbn      = {9780262300414},
	url       = {https://books.google.co.in/books?id=1tbxCwAAQBAJ}
}

@mastersthesis{fatima2025developer,
	author    = {Urooj Fatima},
	title     = {Developer Social Networks / Open-Source Project Networks -- How Programmers Use GitHub},
	school    = {Lappeenranta--Lahti University of Technology LUT},
	address   = {Lappeenranta, Finland},
	year      = {2025},
	url       = {https://urn.fi/URN:NBN:fi-fe2025062472992},
	note      = {In co-operation with partner University: Aalborg University (Copenhagen)}
}

@article{decan2019empirical,
	author    = {Decan, Alexandre and Mens, Tom and Grosjean, Philippe},
	title     = {An empirical comparison of dependency network evolution in seven software packaging ecosystems},
	journal   = {Empirical Software Engineering},
	volume    = {24},
	number    = {1},
	pages     = {381--416},
	year      = {2019},
	publisher = {Springer},
	doi       = {10.1007/s10664-017-9589-y}
}

@inproceedings{linaaker2024sustaining,
	author    = {Lin{\aa}ker, Johan and Link, Georg and Lumbard, Kevin},
	title     = {Sustaining maintenance labor for healthy open source software projects through human infrastructure: A maintainer perspective},
	booktitle = {Proceedings of the 18th ACM/IEEE International Symposium on Empirical Software Engineering and Measurement},
	pages     = {37--48},
	year      = {2024},
	doi       = {10.1145/3674805.3686667}
}

@phdthesis{qi2009comparison,
	author    = {Qi, Jiezhi},
	title     = {Comparison of proportional hazards and accelerated failure time models},
	school    = {University of Saskatchewan},
	year      = {2009},
	url		  = {https://hdl.handle.net/10388/etd-03302009-140638}
}

@article{moran2008modelling,
	author    = {Moran, John L and Bersten, Andrew D and Solomon, Patricia J and Edibam, Cyrus and Hunt, Tamara and Australian and New Zealand Intensive Care Society Clinical Trials Group},
	title     = {Modelling survival in acute severe illness: Cox versus accelerated failure time models},
	journal   = {Journal of Evaluation in Clinical Practice},
	volume    = {14},
	number    = {1},
	pages     = {83--93},
	year      = {2008},
	publisher = {Wiley Online Library},
	doi       = {10.1111/j.1365-2753.2007.00806.x}
}

@article{orbe2002comparing,
	author    = {Orbe, Jesus and Ferreira, Eva and N{\'u}{\~n}ez-Ant{\'o}n, Vicente},
	title     = {Comparing proportional hazards and accelerated failure time models for survival analysis},
	journal   = {Statistics in Medicine},
	volume    = {21},
	number    = {22},
	pages     = {3493--3510},
	year      = {2002},
	publisher = {Wiley Online Library},
	doi       = {10.1002/sim.1251}
}

@article{louis1981nonparametric,
	author    = {Louis, Thomas A},
	title     = {Nonparametric analysis of an accelerated failure time model},
	journal   = {Biometrika},
	volume    = {68},
	number    = {2},
	pages     = {381--390},
	year      = {1981},
	publisher = {Oxford University Press},
	doi       = {10.1093/biomet/68.2.381}
}

@article{patel2006comparing,
	author    = {Patel, Katie and Kay, Richard and Rowell, Lucy},
	title     = {Comparing proportional hazards and accelerated failure time models: an application in influenza},
	journal   = {Pharmaceutical Statistics: The Journal of Applied Statistics in the Pharmaceutical Industry},
	volume    = {5},
	number    = {3},
	pages     = {213--224},
	year      = {2006},
	publisher = {Wiley Online Library},
	doi       = {10.1002/pst.213}
}

@inproceedings{avelino2019abandonment,
	author    = {Avelino, Guilherme and Constantinou, Eleni and Valente, Marco Tulio and Serebrenik, Alexander},
	title     = {On the abandonment and survival of open source projects: An empirical investigation},
	booktitle = {2019 ACM/IEEE International Symposium on Empirical Software Engineering and Measurement (ESEM)},
	pages     = {1--12},
	year      = {2019},
	organization = {IEEE},
	doi       = {10.1109/ESEM.2019.8870181}
}

@article{Evangelopoulos2008,
	author    = {Nicholas Evangelopoulos and Anna Sidorova and Stergios Fotopoulos and Indushobha Chengalur-Smith},
	title     = {Determining Process Death Based on Censored Activity Data},
	journal   = {Communications in Statistics—Simulation and Computation},
	volume    = {37},
	number    = {8},
	pages     = {1647--1662},
	year      = {2008},
	doi       = {10.1080/03610910802140224}
}

@article{Calefato2022,
	author    = {Fabio Calefato and Marco Aurélio Gerosa and Giuseppe Iaffaldano and Filippo Lanubile and Igor Steinmacher},
	title     = {Will you come back to contribute? Investigating the inactivity of OSS core developers in GitHub},
	journal   = {Empirical Software Engineering},
	year      = {2022},
	doi       = {10.1007/s10664-021-10012-6}
}

@article{ohaegbulem2024remedying,
	author    = {Ohaegbulem, Emmanuel Uchenna and Iheaka, Victor Chijindu},
	title     = {On remedying the presence of heteroscedasticity in a multiple linear regression modelling},
	journal   = {African Journal of Mathematics and Statistics Studies},
	volume    = {7},
	number    = {2},
	pages     = {225--261},
	year      = {2024},
	doi       = {10.52589/AJMSS-TJ9XI8HD}
}

@article{karakaplan2020solution,
	author    = {Karakaplan, Mustafa U and Kutlu, Levent and Tsionas, Mike G},
	title     = {A solution to log of dependent variables with negative observations},
	journal   = {Journal of Productivity Analysis},
	volume    = {54},
	number    = {2},
	pages     = {107--119},
	year      = {2020},
	publisher = {Springer},
	doi       = {10.1007/s11123-020-00587-5}
}

@inproceedings{tulili2025exploring,
	author    = {Tulili, Tien Rahayu and Rastogi, Ayushi and Capiluppi, Andrea},
	title     = {Exploring turnover, retention and growth in an OSS Ecosystem},
	booktitle = {Proceedings of the 29th International Conference on Evaluation and Assessment in Software Engineering},
	pages     = {887--897},
	year      = {2025},
	doi       = {10.1145/3756681.3757050}
}

@inproceedings{Dabic2021,
	author    = {Ozren Dabic and Emad Aghajani and Gabriele Bavota},
	title     = {Sampling Projects in GitHub for MSR Studies},
	booktitle = {Proceedings of the 2021 IEEE/ACM 18th International Conference on Mining Software Repositories (MSR 2021)},
	pages     = {560--564},
	year      = {2021},
	publisher = {Institute of Electrical and Electronics Engineers Inc.},
	doi       = {10.1109/MSR52588.2021.00074},
	isbn      = {9781728187105},
	month     = {5}
}

@inproceedings{koch2024fault,
	author    = {Koch, Simon and Klein, David and Johns, Martin},
	title     = {The Fault in Our Stars: An Analysis of GitHub Stars as an Importance Metric for Web Source Code},
	booktitle = {Workshop on Measurements, Attacks, and Defenses for the Web (MADWeb)},
	year      = {2024},
	doi       = {10.14722/madweb.2024.23004}
}

@article{li2021you,
	author    = {Li, Zhixing and Yu, Yue and Wang, Tao and Yin, Gang and Li, Shanshan and Wang, Huaimin},
	title     = {Are you still working on this? An empirical study on pull request abandonment},
	journal   = {IEEE Transactions on Software Engineering},
	volume    = {48},
	number    = {6},
	pages     = {2173--2188},
	year      = {2021},
	publisher = {IEEE},
	doi       = {10.1109/TSE.2021.3053403}
}

@inproceedings{xia2023understanding,
	author    = {Xia, Xiaoya and Zhao, Shengyu and Zhang, Xinran and Lou, Zehua and Wang, Wei and Bi, Fenglin},
	title     = {Understanding the archived projects on GitHub},
	booktitle = {2023 IEEE International Conference on Software Analysis, Evolution and Reengineering (SANER)},
	pages     = {13--24},
	year      = {2023},
	organization = {IEEE},
	doi       = {10.1109/SANER56733.2023.00012}
}

@inproceedings{maldeniya2020herding,
	author    = {Maldeniya, Danaja and Budak, Ceren and Robert Jr, Lionel P and Romero, Daniel M},
	title     = {Herding a deluge of good samaritans: How GitHub projects respond to increased attention},
	booktitle = {Proceedings of The Web Conference 2020},
	pages     = {2055--2065},
	year      = {2020},
	doi       = {10.1145/3366423.3380272}
}

@article{shen2025external,
	author    = {Shen, Yang and Wang, Tao and Zhang, Xunhui and Zhang, Yang and Yang, Cheng and Yu, Yue and Wang, Huaimin},
	title     = {Are External Contributions Important to Project Productivity in Open Source Software? A Deep Insight based on Issue Entropy},
	journal   = {Proceedings of the ACM on Human-Computer Interaction},
	volume    = {9},
	number    = {7},
	pages     = {1--26},
	year      = {2025},
	publisher = {ACM New York, NY, USA},
	doi       = {10.1145/3757399}
}

@inproceedings{puhlfurss2022exploratory,
	author    = {Puhlf{\"u}r{\ss}, Tim and Montgomery, Lloyd and Maalej, Walid},
	title     = {An exploratory study of documentation strategies for product features in popular GitHub projects},
	booktitle = {2022 IEEE International Conference on Software Maintenance and Evolution (ICSME)},
	pages     = {379--383},
	year      = {2022},
	organization = {IEEE},
	doi       = {10.1109/ICSME55016.2022.00043}
}

@inproceedings{tan2023wait,
	author    = {Tan, Wen Siang and Wagner, Markus and Treude, Christoph},
	title     = {Wait, wasn’t that code here before? Detecting Outdated Software Documentation},
	booktitle = {2023 IEEE International Conference on Software Maintenance and Evolution (ICSME)},
	pages     = {553--557},
	year      = {2023},
	organization = {IEEE},
	doi       = {10.1109/ICSME58846.2023.00071}
}

@inproceedings{aghajani2019software,
	author    = {Aghajani, Emad and Nagy, Csaba and Vega-M{\'a}rquez, Olga Lucero and Linares-V{\'a}squez, Mario and Moreno, Laura and Bavota, Gabriele and Lanza, Michele},
	title     = {Software documentation issues unveiled},
	booktitle = {2019 IEEE/ACM 41st International Conference on Software Engineering (ICSE)},
	pages     = {1199--1210},
	year      = {2019},
	organization = {IEEE},
	doi       = {10.1109/ICSE.2019.00122}
}

@inproceedings{dagenais2010creating,
	author    = {Dagenais, Barth{\'e}l{\'e}my and Robillard, Martin P},
	title     = {Creating and evolving developer documentation: understanding the decisions of open source contributors},
	booktitle = {Proceedings of the 18th ACM SIGSOFT International Symposium on Foundations of Software Engineering},
	pages     = {127--136},
	year      = {2010},
	doi       = {10.1145/1882291.1882312}
}

@inproceedings{Aggarwal2014,
	author    = {Karan Aggarwal and Abram Hindle and Eleni Stroulia},
	title     = {Co-evolution of project documentation and popularity within GitHub},
	booktitle = {Proceedings of the 11th Working Conference on Mining Software Repositories (MSR 2014)},
	pages     = {360--363},
	year      = {2014},
	publisher = {Association for Computing Machinery},
	doi       = {10.1145/2597073.2597120},
	isbn      = {9781450328630},
	month     = {5}
}

@article{jiang2017,
	author    = {Jiang, Jing and Lo, David and He, Jiahuan and Xia, Xin and Kochhar, Pavneet Singh and Zhang, Li},
	title     = {Why and how developers fork what from whom in GitHub},
	journal   = {Empirical Software Engineering},
	volume    = {22},
	number    = {1},
	pages     = {547--578},
	year      = {2017},
	publisher = {Springer},
	doi       = {10.1007/s10664-016-9436-6}
}

@inproceedings{steinmacher2015social,
	author    = {Steinmacher, Igor and Conte, Tayana and Gerosa, Marco Aur{\'e}lio and Redmiles, David},
	title     = {Social barriers faced by newcomers placing their first contribution in open source software projects},
	booktitle = {Proceedings of the 18th ACM Conference on Computer Supported Cooperative Work \& Social Computing},
	pages     = {1379--1392},
	year      = {2015},
	doi       = {10.1145/2675133.2675215}
}

@article{Kaushik2026BeyondSpeed,
	author    = {Kaushik, Mohit and Chahal, Kuljit Kaur},
	title     = {Beyond Speed: Engagement Sustains Lifespan},
	journal   = {Software: Practice and Experience},
	volume    = {56},
	number    = {6},
	pages     = {758--785},
	year      = {2026},
	publisher = {Wiley},
	doi       = {10.1002/spe.70068}
}

@inproceedings{steinmacher2014barriers,
	author    = {Steinmacher, Igor and Silva, Marco Aur{\'e}lio Graciotto and Gerosa, Marco Aur{\'e}lio},
	title     = {Barriers faced by newcomers to open source projects: a systematic review},
	booktitle = {IFIP International Conference on Open Source Systems},
	pages     = {153--163},
	year      = {2014},
	organization = {Springer},
	doi       = {10.1007/978-3-642-55128-4_21}
}

@article{buse2010readability,
	author    = {Raymond P. L. Buse and Westley R. Weimer},
	title     = {Learning a Metric for Code Readability},
	journal   = {IEEE Transactions on Software Engineering},
	volume    = {36},
	number    = {4},
	pages     = {546--558},
	year      = {2010},
	doi       = {10.1109/TSE.2009.70}
}

@article{Wang2012,
	author    = {Jing Wang},
	title     = {Survival factors for Free Open Source Software projects: A multi-stage perspective},
	journal   = {European Management Journal},
	volume    = {30},
	number    = {4},
	pages     = {352--371},
	year      = {2012},
	publisher = {Elsevier},
	doi       = {10.1016/j.emj.2012.03.001}
}

@article{Kaushik2025,
	author    = {Mohit Kaushik and Kuljit Kaur Chahal},
	title     = {Community Engagement and the Lifespan of Open-Source Software Projects},
	journal   = {Information and Software Technology},
	volume    = {189},
	pages     = {107914},
	year      = {2026},
	doi       = {10.1016/j.infsof.2025.107914},
	url       = {https://doi.org/10.1016/j.infsof.2025.107914}
}

@article{Joblin2022,
	author    = {Mitchell Joblin and Sven Apel},
	title     = {How Do Successful and Failed Projects Differ? A Socio-Technical Analysis},
	journal   = {ACM Transactions on Software Engineering and Methodology},
	volume    = {31},
	number    = {4},
	pages     = {67:1--67:24},
	year      = {2022},
	publisher = {Association for Computing Machinery},
	doi       = {10.1145/3504003}
}

@book{eghbal2020working,
	author    = {Eghbal, Nadia},
	title     = {Working in Public: The Making and Maintenance of Open Source Software},
	year      = {2020},
	publisher = {Stripe Press},
	address   = {San Francisco, California},
	isbn      = {978-0-578-67586-2},
	url       = {https://books.google.co.in/books?id=zxjBEAAAQBAJ}
}

@article{Kaushik2026DeathSpiral,
	author    = {Mohit Kaushik and Kuljit Kaur Chahal},
	title     = {The death spiral of open source projects: A post-mortem analysis of pull request workflow dynamics},
	journal   = {Journal of Systems and Software},
	volume    = {240},
	pages     = {112942},
	year      = {2026},
	publisher = {Elsevier},
	doi       = {10.1016/j.jss.2026.112942}
}
\end{document}